\documentclass[fleqn,twoside,11pt]{article}
\pdfoutput=1
\usepackage{amsfonts,amsmath,amssymb}
\usepackage{latexsym}
\usepackage[utf8]{inputenc}
\usepackage[english,russian]{babel}
\usepackage{verbatim}
\usepackage{cite}

\begin{document}

\twocolumn[
\title{{\bf On Geometrization of Classical Fields\\  (Model of Embedded Spaces)}}
\author{{\bf V. I. Noskov}}	       	
\date{\small \em Institute of Continuum Media Mechanics, Ural Branch of the Russian Academy of Sciences, Perm, Russia,
Email: nskv@icmm.ru}

\maketitle
{\bf Abstract --}{\small
The possibility of geometrization of the gravitational and electro magnetic fields in 4D Finsler space (the Model of Embedded Spaces -- MES) is investigated. The model postulates a proper metric set of an
{\em element} of distributed matter and asserts that space-time is a mutual physical {\em embedding} of such sets. The simplest  MES geometry is constructed (its {\em relativistic} Finsler version) with
a connection that depends of the properties of matter and its fields (torsion and nonmetricity are absent). The field hypothesis and the Least Action Principle of the matter-field system lead to Einstein-type
and Maxwell-type equations, and their nonlinearity -- to the {\em anisotropic} field contribution to the seed mass of matter. It is shown that the seed matter plays the role of a physical vacuum of the
{\em Embedding} determines the cosmological constant. In the special case of a conformal metric, the Maxwell-type equations reduce to the Maxwell equations themselves and a negative electromagnetic
contribution. A possible experimental verification of this result is evaluated. The "redshift" effect in an electric field is also mentioned as a method for studying the vacuum and relic electric charge. A study
of the gauge structure of the presented theory is postponed to the future.}
\medskip

{\bf DOI:} 10.1134./S0202289323020081
\bigskip
]

\section{\textmd{INTRODUCTION}}
\label{intro}
\quad The necessity of studying the geometrical aspect of physical phenomena was clearly recognised long before Descartes, and with the invention of the coordinate system, the progress in this area was significantly
accelerated. A remarkable example of demonstrating the role of geometry in physics is E. Noether's famous theorem (1918) on the geometric and symmetry meaning of physical conservations laws. Bat a real breakthrough
in the study of the physical role of the geometry of space was the classical works of A. Einstein and a large group of his followers in the first half of the 20th century, devoted to the creation of special and general relativity \cite{Ein1905,Ein1916}. The ideological simplicity and effectiveness of these works caused an explosion of interest not only in geometrization of gravity and electromagnetism, but also in geo\-metrization of physics as a
whole. The relativistic pseudo-Euclidian \cite{Minkovski1909} and later pseudo-Riemann 4-dimen\-sional manifolds \cite{Nordstrom1913,Ein1915} became physical stan\-dards. There was a conviction that prospects for
the development of physics in this direction are very promising, and it is necessary to follow it. In particular, in addition to finding out the universal properties of physical space, there loomed a possi\-bility of combining
classical fields and matter on a single geometric basis. At the same time, there seemed to be many options in choosing such a basis: from various special ({\em configuration}) cases of Riemann geometry (conformity,
extra dimensions, torsion, etc.) to its anisotropic genera\-lization, the Finsler geometry, invented in 1918 and intensively developed till the middle of the 20th century \cite{Rund1959}.

However, time has decided otherwise. The discovery and rapid development of quantum me\-chanics, the observation of previously uncnown strong and weak interactions, and the interpretati\-on of "old"\ physics as the
classical limit of quantum physics have led to a shift in the research priorities. But the work on geometrization of physics, although not with the same intensity, still continued, based mainly on Riemann (it gradually
became clear that only a partially suitable one) geometry, in the "early"\ \cite{EinFocker1914,Weil1918,Kaluza1921,Klein1926,Fokker1929,Pauli1933,Edd1943,Edd1946} and "later"\ \cite{Rumer1956,Zimmerman1962,Rumer1971,Penrose1972,Konopl1972,Schmutzer1980,Salam1982,Vlad1982,Kokarev1995,Krechet1995,Wesson1996} studies. There were also papers that used the very promising
Finsler \cite{Rund1959} anisotropic model of space, for example, \cite{Voicu2011}. However, the generally recognized success in the theory was never achieved, although a preliminary and very important step in
using the Cartan-Finsler geometric idea was nevertheless made the papers by G. Yu. Bogoslovsky and co-authors \cite{Bo1973,Bo1992,Bo1996,Bo2012}. The relativistic metric studied there, going beyond the
framework of standard Finsler geometry(see Section 4 and App\-endix B), gives a necessary anisotropic basis for a {\em relativistic} Finsler geometry, similar to the Minkow\-ski metric for the pseudo-Riemann
geometry of GR.

Against this background, the works of the group of authors led by Yu. S. Vladimirov  \cite{Vlad1996,Vlad1998,Vlad2009} also related to geometrization of physics, but {\em in the paradigm of direct particle interaction},
are also quite noticeable, but are only indirectly related to the further presentation.

The presently relevant physical problems, such as quantization of gravity, a description of the evolution of the observable universe, progress in the extremely complex string theory, etc., related to the geometry of space-
time on almost all scales, permanently force us to return to the question of its real structure. An example is the crisis of the standard cosmological model caused by recent astronomical discoveries \cite{Riess1998,Perlmut1999}.
It suggests that the electrically neutral model of GR space is incomplete, ar least on cosmological scales \cite{Rubakov2007}. At the same time, it seems obvious that the inclusion of electromagnetism in the geometry of
the Standard Model can both significantly change its evolutionary behavior and lead to the appearance of properties radically different from the standard ones. For example, due to a nonzero {\em relic} electric charge of the
Universe.

The above is quite sufficient for returning to the "eternal"\ problem of geometrization of classical electricity.

At the same time, it seems natural that the correct solution of the latter problem should be sought within the framework of a different, non-Einstein-Riemann, obviously non-configuration model of space-time. The geodesic
equation of the desired model must contain, in addition to the usual gravitational one, also a Lorentzian term that depends on the {\em inherent} characteristics of matter (charge, mass, velocity), which gives it an explicit
{\em configuration} meaning of the old model! In other words, the new model of space-time should naturally remove  this configuration contradiction. In addition, it should "work"\ for classical fields and matter on scales
ranging at least from molecular to cosmological ones. That is, the model space must  be smooth and 4-dimensional (have a tangent Minkowski space \cite{Minkovski1909} at each point) due to the experimental non-observability
of higher dimensions. Second, the model must be {\em metric}. This restriction is significantly less experimentally reliable than the first one (in terms of small scales), but the success of GR in describing gravity seems to be a serious argument in favor of its validity. And third, in the limit of only the gravita\-tional interaction, the desired model should lead to GR.

An almost successful implementation of the formulated requirements is the mentioned Finsler model (a fiber bundle with the velocity field of matter as the tangent fiber \cite{Rund1959}), equipped with the corresponding
geometry. At the first glance, the model has all necessary properties for the simplest geometrization of electrodynamics and combining it with gravity: Finsler's geometry is metric, local, and anisotropic -- the metric tensor
depends both on the coordinates $x$ and on the direction $\dot{x}$ (the tangent vector of any curve passing through $x$). (The anisotropy is necessary because Maxwell's electrodynamics is a vector theory, defined by
the 4-vector $A_i=(\varphi_e, -\vec A)$.) Why was the Finsler model nonproductive for the geometri\-zation and unification?

The answer is simple: the standard Finsler geometry is {\em non-relativistic} \cite{Nskv2001,Nskv2004-1,Nskv2004-2}. It allows for parametrization by {\em only} non-geometric Newto\-nian time (1D-extended
geometric quantities like $x^0$ are forbidden), and the geodesic {\em cannot} have a Lorentzian term. A study of this aspect of geometry with the corresponding proofs is presented in Appendix A. In addition, the
{\em non-configurationa\-lity} of the Finsler model is simply postulated, and the physical reasons for the $\dot{x}$-dependence of the metric are unclear: there is no {\em physical} model of the Finsler bundle.

Thus the Finsler tangent bundle equipped with a relativistic version of the Finsler geometry can still be taken as the desired mathematical model of space-time. Obviously, to illustrate the physical properties of the
mathematical model and its geometry, a suitable physical non-configura\-tion space-time model is needed. The simplest of these is the MES of \cite{Nskv2007,Nskv2013}. It assumes that any element of distributed
matter has its own smooth metric set, and the real 4D space-time of the Universe is a mutual {\em physical embedding} of these sets into each other. The interaction of elements plays the role of a "glue"\ holding
 the {\em Embedding} from breaking up into proper sets.

(It is obvious that the proper sets of matter ensure the {\em non-configurationality} of the model, and the {\em "Embedding"\ } manyfold is metric and is characterized by the internal relative motion of the original
sets. An estimate of the size of the partial set can be taken from \cite{Edd1943}, p. 12: "...from $10^4\, mm$ to $10^7\,pc$"\ . The usage of the term {\em "Embedding"\ } without appropriate mathema\-tical
reservations and clarifications is not entirely correct, but the "physical"\ definition successfully expresses the essence of the matter. For clarify: in what follows, {\em Embedding} can be understood as a total point
set of continuum cardinality, which is analogous to continuous medium, for example, {\em mixture} of various fluids. It is important that all partial sets equally form an {\em Embedding}.)

The initial effort in the development of MES are available in \cite{Nskv2001,Nskv2004-1,Nskv2004-2,Nskv2007,Nskv2008}, as well as fragmentarily in the recent papers \cite{Nskv2013,Nskv2016,Nskv2017}. In this
paper the basics of the relativistic Finsler geometry of the {\em Embedding}, the geometrization of electricity and gravity on its basis are consistently presented. Some new and previously discovered physical and
cosmological consequences of geomet\-rization are presented and discussed.

\section{\textmd{{\em EMBEDDING} IN THE TEST\\ PARTICLE APPROXIMATION}}
\label{DinEm}
\quad Obviously, the material base of the MES is a relativistic classical (non-quantum) electrically charged continuous medium filling the Universe. A consistent description of such a medium is possible only in two cases:
a medium consisting of interacting material points, and a medium whose properties are completely determined by the field hypothesis of matter. Without specifying , further on we will proceed from the assumption that
there is a continuous electrically charged material medium.

The simplest version of the MES \cite{Nskv2013} ({\em Embed\-ding}) uses the approximation of a test particle  and assumes the existence of only two initial sets $M_c$ (charge) -- the set of test element of electrically
charged distributed matter, and $M_s$ (source) -- the set of a field source to which belongs the whole other matter in the Universe. In this approximation, the {\em Embedding} metric, like the Finsler case, is also
determined by the local mecha\-nical state of matter,
\begin{equation}
\label{1}
g_{ik}=g_{ik}(x^l,u^m),
\end{equation}
where $x^l$ are 4-coordinates,  $u^m=dx^m/ds$ is the 4-velocity of matter at the point under considera\-tion in an arbitrary coordinate system of the {\em Embedding}, and $ds=+\sqrt{g_{ik}dx^idx^k}$ is its
interval,\footnote{The dimension of the original sets is not yet specified -- this is a subject for a separate study. Although it seems reasonable to choose 3D (only spatial dimensions), and the dimension $x^0$
of the {\em Embedding} is a natural consequence of the mobility of the original sets.} and for weak fields (small velocities of the test matter), \eqref{1} looks as
\begin{eqnarray}
 \label{2}
 \nonumber
   g_{00}\simeq 1+2\left(\varphi_g+\beta Au\right)/c^2,\\
   g_{0\alpha}\simeq g^{(0)}_{0\alpha}=0, \quad  g_{\alpha\beta}\simeq g^{(0)}_{\alpha\beta}=-\delta_{\alpha\beta},
\end{eqnarray}
where $Au\simeq \varphi_e-\vec A\,\vec v/c$ and $\varphi_g$, $\varphi_e$, $\vec A)$ are the Newtonian and Maxwellian potentials, $\beta=\rho/\mu$ is the ratio of the electric charge and mass densities.

{\bf Explanation.}

1. $dx^l$ is a chord of a trajectory passing through the points $x^l$ and $x^l+dx^l$, \cite{Rund1959}, chapter 1, \$1.

2. The choice of the natural parameter $s$ guarantees the relativistic nature and unit length of tangent vector $u$ in the {\em Embedding}. In this case, Eq. \eqref{1} becomes an implicit definition of the metric $g_{ik}$,
which does not restrict the latter in any way.

3. $A_i$ in \eqref{2} is a potential with the gauge $A_i \rightarrow A_i+\partial f/\partial x^i$, where (a) in the tangent (Minkowski) space $f=f(x)$, and the compact group $U(1)$ as a consequence of the derivation
\eqref{2} ($[\mu c^2\sqrt{1-\vec v^2/c^2}+\mu \varphi_g+\rho(\varphi_e-\vec A\cdot\vec v/c)]dt=\mu c\, ds$, see the end of \$87 in \cite{LL1988}) and (b) in the {\em Embedding}, $f=f(x,u)$, (anisotropic)
gauge group is still unknown, see also an Explanation at the end of Subsection \ref{CovDer}.

The form of the metric \eqref{2} indicates that it belongs to the class of Finslerian {\em inhomogeneous} metrics (Appendiix A): is not a {\em homogeneous} function of degree $0$ of $u^i$.  That is , the standard
Finsler ({\em homogeneous}) geometry \cite{Rund1959}  {\em does not des\-cribe} the {\em Embedding} (as well as the Bogoslovsky space, see (Appendix A)).

Thus, for a further study, it is necessary to develop a {\em inhomogeneous} (relativistic) version of Finsler geometry, which will simultaneously geometrize gravity and electricity. As will be shown, such a geometry
can be constructed in the image and likeness of Riemann geometry, since the aniso\-tropic argument $u^m$ of the metric tensor \eqref{1} can be considered as a field of directions $u^m(x)$ (the Eulerian description
of a continuous medium). And a considering smooth coordinate transforma\-tions $x^i=x^i(x^{'k},u^{'l})$ of the {\em Embedding} $M_e$ refe\-rence frames and their Jacobians, we can construct a generally covariant
anisotropic tensor analysis for the new geometry by the analogy with the Riemann differential formalism: in the usual way, we define the contravariant tensor $g^{ik}(x^l,u^m)$ as the inverse of the tensor
\eqref{1}, such that $g_{ik}g^{kl}=g_i^{\ l}\equiv\delta_i^{\ l}$, using the standard rules for lifting the tensor indices, the connectedness, etc., preserve the norms $g_{ik}g^{ik}=4$ and
$g_{ik}u^iu^k\equiv u_iu^i=1$, and so on.

It should also be noted that a formal generali\-zation is possible, which leads to a pure geometry model: the pare $(x^l,u^m)$ in \eqref{1} can be understood as the state of arbitrary open curve passing through
the point $x$ of the {\em Embedding}.

\subsection{\textmd{Geodesics (Equation of Motion)}}
\label{Geod}
\quad As is known, in geometrized physical theory, matter moves by inertia, along geodesic curves of the space-time geometry. That is, the geodesic equation in {\em Embedding} is also an equation of motion
of matter in the considered fields (gravita\-tional and electromagnetic ones).

We use the standard variational method, assu\-ming that the geodesic is the shortest line connect\-ing the points $x_1$ and $x_2$:
\begin{equation*}
\delta\int_{x_1}^{x_2}ds=0,
\end{equation*}
where $ds=\sqrt{g_{ik}(x,u)dx^idx^k}$ is the interval of the curve, $u=dx/ds$ is a unit tangent vector, and $s$ is a natural parameter (lenth).

Then, for variation of the interval we can write
\begin{equation*}
0=\delta\sqrt{g_{ik}(x,u)dx^idx^k}=u_i\delta dx^i+\frac{dx^kdx^l}{2ds}\cdot
\end{equation*}
\begin{equation*}
\cdot\left[\frac{\partial g_{kl}}{\partial x^i}\delta x^i+\frac{\partial g_{kl}}{\partial u^m}
\left(\frac{\delta dx^m}{ds}-u^m\frac{\delta_x ds}{ds}\right)\right]=
\end{equation*}
where $\delta x^i$ is the variation vector (equal to zero at the ends of the curve), and $\delta_x$ is the variation operator for only $x$- dependence in $ds(x,u)$ (rest\-ricted to the first iteration -- linear approximation).

Using the commutativity of operators $\delta$ and $d$ and omitting the full differentials of the first and last two terms, continuing the equality, we obtain
\begin{equation*}
  =ds\delta x^i\left[-\frac{du_i}{ds}+u^ku^l\left(\frac12\frac{\partial g_{kl}}{\partial x^i}\right)\right]+\vartheta_i\left(d\delta x^i-\right.
\end{equation*}
\begin{equation*}
   \left.-u^id\delta_x s\right)=ds\left\{ \delta x^i\left[\cdots\right]-\delta x^i\frac{d\vartheta_i}{ds}+\delta_x s\frac{d(\vartheta u)}{ds}\right \},
\end{equation*}
vhere $\vartheta_i$ is the vector quantity\footnote{It has three independent components (as the potential $A_i$) due to the three independent components of the vector $u^i$.}
\begin{equation}
 \label{3}
  \vartheta_i=c_{i,kl}u^ku^l,\qquad 2c_{i,kl}=\partial g_{kl}/\partial u^i,
\end{equation}
and $c_{i,kl}$ is a tensor one (recall that $u^i=u^i(s)$).

Using the Eulerian form $d/ds\equiv u^i\partial/\partial x^i$, the operator of the last term in the curly brackets can be transformed as
\begin{equation*}
  \delta_xs\,d/ds=\delta_xs\,u^i\partial/\partial x^i\simeq \delta x^i\,\partial/\partial x^i
\end{equation*}
and putting the velocities out of the coordinate derivatives, we finally have
\begin{equation}\label{4}
  \frac{du_i}{ds}-\frac12u^ku^l\left(\frac{\partial}{\partial x^i}+2u^m\frac{\partial^2}{\partial x^{[i}\partial u^{m]}}\right)g_{kl}=0,
\end{equation}
where the square brackets at the indices indicate antisymmetrization,
\begin{equation*}
  \partial^2/\partial x^{[i}\partial u^{k]}=\left(\partial^2/\partial x^i\partial u^k-\partial^2/\partial x^k\partial u^i\right)/2.
\end{equation*}

There important conclusions follow from \eqref{4}:

First, it is clear that the last term ia Lorentzi\-an: the geometrization performed leads to a tensor version of electrodynamics with the potential $c_{i,kl}$ \eqref{3}, more general than the classical one. A connec\-tion  with
Maxwell's vector electrodynamics is imple\-mented by the definition of the vector $\vartheta_i$;

Second, it is to prove that the second term in \eqref{4} is purely gravitational: the derivative $\partial g_{kl}/\partial x^i$ is taken at $u^m=const$, that is, $\partial g_{kl}(x, u=const)/\partial x^i$. And under this
condition, the electro\-magnetic "potentials"\ \eqref{3} are identical zeros: \\ $2c_{i,kl}=\partial g_{kl}(x, u=const)/\partial u^l\equiv 0 \Rightarrow \vartheta_i\equiv 0$. Hence, the second term in \eqref{4} is related
{\em to gravity only}: $\partial g_{kl}/\partial x^i\sim \partial \varphi_g/\partial x^i$, $\Box$;

Third, as one could expect, the continuous medium filling the {\em Embedding} moves along its geodesics (see also Appendix C).

\subsection{\textmd{The Covariant Derivative}}
\label{CovDer}
\quad The form of the geodesic \eqref{4} and the structure of {\em Embedding} $M_e$ (sets $M_s$ and $M_c$, the stationary one and the one moving with velocity $u^i$) indicate the fact that in differential
analysis of the MES geometry there are {\em two different gradient operators} $\partial/\partial x^i$ and $\hat b_i$,
\begin{equation}\label{5n}
  \partial/\partial x^i\in M_s,\quad \hat b_i\in M_c,
\end{equation}
discribing the differential increment of an arbitrary $f(x,u)$. Outside the sets, they do not exist ({\em are zero}):
\begin{equation}\label{5nn}
  \partial/\partial x^i=0\notin M_s,\quad \hat b_i=0\notin M_c.
\end{equation}

Obviously, the gradients determine the main contributions due to $M_s$ and $M_c$ in the {\em Embedding},
\begin{equation*}
 \bar d=dx^i\:\bar\partial/\partial x^i, \quad \bar\partial/\partial x^i=\partial/\partial x^i+\hat b_i,
\end{equation*}
\begin{equation}\label{5}
 \hat b_i=2u^k\hat b_{ik},\quad \hat b_{ik}=\partial^2/\partial x^{[i}\partial u^{k]},
\end{equation}
where  $\hat b_{ik}=-\hat b_{ki}$. The operators $\hat b_i$ and $\hat b_{ik}$ are the {\em projection, linear, first-order} operators since they give a first-order increment in $dx^i$\footnote{Example:
$\hat b_{ik}(1+c^{-2}\beta Au)=c^{-2}(\beta\hat b_{ik}Au+Au\hat b_{ik}\beta)=c^{-2}\beta\hat b_{ik}Au=2c^{-2}\partial A_{[k}/\partial x^{i]}$.}. Obviously, the Lorentz gradient $\hat b_i$ is
directly accounting for the anisotropic (the set $M_c$) pro\-perties of the {\em Embedding}, and $\partial/\partial x^i$ for the isotro\-pic ones (the set $M_s$).

The metricity of the manifold $M_e$ and the original sets $M_s$ and $M_c$ requires assumptions on the corresponding geometries: {\em we assume that there are no non-metricity and torsion, and every\-thing
reduces to connections}.  And while the geome\-try of $M_s$ is clearly Riemann (GR), then, {\em assuming} the other geometries to be Riemann, we can const\-ruct the corresponding differential formalism by
introducing covariant generalizations of the paral\-lel transfer and coordinate derivatives \cite{Nskv2013}:
\begin{equation}
 \label{8}
  \ldots_{|i}\equiv \bar \partial\ldots /\partial x^i\pm\Gamma^._{\ i.}\ldots,\quad g_{ik|l}=0
\end{equation}
for $M$ ($\Gamma_{i,kl}\in M$) and
\begin{equation}\label{9}
  \ldots_{|i}\equiv \ldots_{;i}+ \ldots_{:i},
\end{equation}
where
\begin{equation*}
  \ldots_{;i}\equiv \partial\ldots /\partial x^i\pm\gamma^._{\ i.}\ldots,
\end{equation*}
\begin{equation}\label{10}
  \ldots_{:i}\equiv \hat b_i\ldots \pm\omega^._{\ i.}\ldots
\end{equation}
for $M_s$ and $M_c$ respectively: $\gamma_{i,kl}\in M_s$, $\omega_{i,kl}\in M_c$.

The simplest interpretation of this generaliza\-tion is reduced to one-to-one correspondence of parallel transfer of the metric
\begin{equation}\label{11}
  g_{ik|l}=0\quad \Leftrightarrow \quad g_{ik;l}=0,\quad g_{ik:l}=0,
\end{equation}
and the additivity of connections
\begin{equation}\label{12}
  \Gamma^i\,_{kl}=\gamma^i\,_{kl}+\omega^i\,_{kl}
\end{equation}
with the explicit expressions for the later
\begin{equation*}
  2\Gamma_{i,kl}=\bar\partial g_{ik}/\partial x^l+\bar\partial g_{il}/\partial x^k-\bar\partial g_{kl}/\partial x^i,
\end{equation*}
\begin{equation*}
  2\gamma_{i,kl}=\partial g_{ik}/\partial x^l+\partial g_{il}/\partial x^k-\partial g_{kl}/\partial x^i,
\end{equation*}
\begin{equation}\label{13}
  2\omega_{i,kl}=\hat b_l g_{ik}+\hat b_k g_{il}-\hat b_i g_{kl}.
\end{equation}

It is quite clear that the domain of existence of any connection coincides with that of its gradient \eqref{5n}, \eqref{5nn}: in particular, any product of the connecti\-ons $\gamma_{i,kl}$ and $\omega_{i,kl}$ is
identically zero. In addition, both simple and generalized mixed derivatives of any function are also zero. The MES has only an {\em Embedding} of sets and no their {\em Intersection}.

Finally, an important corollary should be emp\-hasized: $\gamma_{i,kl}$ contains {\em only gravitational} terms, just as $\omega_{i,kl}$ {\em only Lorentzian} terms. In other words, $\gamma_{i,kl}$ completely
equivalent to the Riemann connec\-tion of GR, it has no "cross"\ terms like
$$\partial (c^{-2}\beta Au)/\partial x^l$$
\eqref{2} (see the end of the Subsection 2.1).

Equation \eqref{8} -- \eqref{13} make it possible to give the geodesic equation \eqref{4} a typical Riemann form:
\begin{equation}\label{16}
 \begin{cases}
  du^i/ds+\Gamma^i_{\ kl}u^ku^l=0, \\
  du_i/ds-\Gamma^k_{\ il}u_ku^l=0.
 \end{cases}
\end{equation}

To demonstrate the Lorentzian summand of the geodesic, it is convenient to use the operators $\hat b_{ik}$ \eqref{5} in the expressions \eqref{13} for the connections $\omega_{i,kl}$
\begin{eqnarray}
\nonumber
\bar Du^i/ds\equiv du^i/ds+\Gamma^i_{\ kl}u^ku^l=\\
\label{17}
  =du^i/ds+\gamma^i_{\ kl}u^ku^l-f^i_{\ k,lm}u^ku^lu^m=0,
\end{eqnarray}
which is a sum of the gravitational and Lorentzian accelerations, and the connection itself $\omega_{i,kl}$ is
\begin{equation}\label{18}
  \omega_{i,kl}=\left(f_{lm,ik}+f_{km,il}-f_{im,kl}\right)u^m,
\end{equation}
where $f_{ik,lm}$ is the "material"\ MES-analogue of the electromagnetic field tensor,
\begin{equation}\label{19}
  f_{ik,lm}\equiv\hat b_{ik}g_{lm}=\partial c_{k,lm}/\partial x^i-\partial c_{i,lm}/\partial x^k
\end{equation}
and $c_{i,kl}$ \eqref{3} is the "material"\ MES-analogue of the vector potential $A_i$.

Finally, we need to give some simple formulas for the connection $\omega_{i,kl}$. The  linearity of $\bar d$ leads to a generalization of the standard relation between the differentials of the determinant and the
compo\-nents of the metric tensor,
\begin{equation*}
  \bar dg=gg^{ik}\bar dg_{ik}=-gg_{ik}\bar dg^{ik}\quad\Rightarrow
\end{equation*}
\begin{equation}\label{20}
  \quad \frac{\partial g}{\partial x^i}=gg^{kl}\frac{\partial g_{kl}}{\partial x^i},\quad \hat b_i g=gg^{kl}\hat b_i g_{kl},\quad \cdots
\end{equation}
whence follow various useful relationships:
\begin{equation*}
  \omega^k_{\ ik}=\hat b_i \ln\sqrt{-g}=f_{il,k}\,^ku^l\quad \Rightarrow
\end{equation*}
\begin{equation}\label{21}
  \quad f_{ik,l}\,^l=2\hat b_{ik}\ln\sqrt{-g};
\end{equation}
\begin{equation*}
  \omega^{i,k}_{\ \ \ k}=-\frac{\hat b_k\left(\sqrt{-g}\,g^{ik}\right)}{\sqrt{-g}}=\left(2f_{kl,}^{\ \ \ ik}-f^{i\ \ \ k}_{\ l,k}\right)u^l\Rightarrow
\end{equation*}
\begin{equation*}
   \quad 2f_{lk,}^{\ \ \ il}-f^{i\ \ \ l}_{\ k,l}=2\hat b_{kl}\left(\sqrt{-g}\,g^{il}\right)/\sqrt{-g}\Rightarrow
\end{equation*}
\begin{equation}\label{22}
  \quad f_{il,k}^{\ \ \ l}= g^{lm}\hat b_{il}g_{km},
\end{equation}
and also
\begin{equation*}
  2\omega_{(i,k)l}=\hat b_lg_{ik},\quad 2\omega^{(i,k)}_{\ \ \ \ l}=-\hat b_lg^{ik},
\end{equation*}
\begin{equation}\label{23}
 \hat b_{ik}g^{lm}=-f_{ik,}^{\ \ \ lm},\quad\omega^{[i,k]}_{\ \ \ \ k}=2f_{l\ \ \ \ \  k}^{\ [i,k]}u^l,\: \cdots
\end{equation}
where the parentheses around a pair of indices denote symmetrization.

An important fact of the MES differential for\-malism is the {\em non-commutativity} of the operators $\hat b_i$:
\begin{equation}\label{24}
  \qquad \hat b_{[i}\hat b_{k]}=d\hat b_{ik}/ds.
\end{equation}

{\bf Explanation.}

1. Obviously, the gauge transformation of the matter-field tensor "potential"\ $c_{i,kl}\rightarrow c_{i,kl}+\partial \Phi_{kl}(x)/\partial x^i$, where $\Phi_{kl}(x)$  is an arbitrary symmet\-ric tensor, is a smooth function
of the point $x$: it does not  change ,$f_{ik,lm}$, nor the connection $\omega_{i,kl}$, nor the curvature $r_{iklm}$. The gauge is based on the corresponding symmetry group, which determines the properties of
"electrodynamics"\ with the "po\-tential"\ $c_{i,kl}$. Identification of the group is not  an easy task: the linear gauge increments of $c_{i,kl}$ are determined by the noncommutative gradients $\hat b_i$ \eqref{5} and
\eqref{24}, and the {\em Embedding} itself is a nontrivial {\em anisotropic} space.

2. The problem of compactification of the gauge group must be reduced to $\omega_{i,kl}$ by virtue of the {\em non-configurational} structure of the {\em Embedding}: it is {\em not} a fiber bundle but contains the
proper space of matter $M_c$.

3. An answer to questions 1 and 2 will allow for qualifying the geometrization in its electromagne\-tic part: does it really lead to nonlinear electrody\-namics?

\subsection{\textmd{Coordinate Transformations}}
\label{CooTrans}
\quad Since the geometry of {\em Embedding} $M_e$ is deter\-mined by the local state of the physical system $(x^i,u^k)$,  the qualification of an arbitrary geometric object is determined by its behavior under local
state transformations $(x^i,u^k)\leftrightarrow (x^{'l},u^{'m})$ of the transfer curve. This leads to replacing arbitrary smooth isotropic coordinate transformations $x^i=x^i(x^{'k})$ of Riemann geometry with
anisotropic ones of Cartan-Finsler type \cite{Rund1959}
\begin{equation}\label{25}
x^i=x^i(x^{'l},u^{'m}),\quad u^k=u^k(x^{'l},u^{'m}),
\end{equation}
where $(x,u)$ are independent variables.

Let us find an expression for the {\em anisotropic} differential of the "old"\ coordinates
\begin{equation}\label{26}
\bar dx^i=\frac{\partial x^i}{\partial x^{'k}}\ dx^{'k}+\frac{\partial x^i}{\partial u^{'k}}\ du^{'k}.
\end{equation}
To do that, consider the second term separately, recalling that $u^{'k}=dx^{'k}/ds'$ and re-denoting for convenience $du^{'k}\equiv \delta u^{'k}$:
\begin{equation*}
\frac{\partial x^i}{\partial u^{'k}}\ du^{'k}=\frac{\partial x^i}{\partial u^{'k}}\left(\frac{\delta (dx^{'k})}{ds'}-\frac{dx^{'k}}{ds^{'2}}\delta ds'\right)=
\end{equation*}
\begin{equation*}
=\frac{d}{ds'}\left[\frac{\partial x^i}{\partial u^{'k}}\left(\delta x^{'k}-u^{'k}\delta s'\right)\right]-\delta x^{'k}\frac{d}{ds'}\left(\frac{\partial x^i}{\partial u^{'k}}\right)+
\end{equation*}
\begin{equation*}
+\delta s'\frac{d}{ds'}\left(u^{'k}\frac{\partial x^i}{\partial u^{'k}}\right)=
\end{equation*}
It can be seen that the term in square brackets in the cord approximation is zero: $\delta x^{'k}-u^{'k}\delta s'=0$, and the operator $d/ds'$ in the last two terms should be presented in the Eulerian form
$d/ds'=u^{'i}\partial/\partial x^{'i}$. Then the independence of the pair of  "new"\ variables, $\partial u^{'i}/\partial x^{'k}=0$, and return to the initial notation of the differential $\delta \rightarrow d$ give,
continuing the previous equality,
\begin{equation*}
=2\delta x^{'l}u^{'k}\frac{\partial^2 x^i}{\partial x^{'[l}\partial u^{'k]}}\equiv dx^{'l} \hat b'_{\ l} x^i.
\end{equation*}

Substituting this result to \eqref{26}, we get
\begin{equation}\label{27}
\bar dx^i=\frac{\bar \partial x^i}{\partial x^{'k}}\ dx^{'k}=dx^i+dx^{'k}\hat b'_{\ k}x^i,
\end{equation}
which is useful to compare with \eqref{5}. It can also be noted that the anisotropic differential $\bar dx^i$ is a function of the local state $(x,u)$, and it is an infinitesimal anisotropic vector in the {\em Embedding} $M_e$
similar to isotropic $dx^i$ in Riemann space. Therefore, the relation between $\bar dx^i$ and $\bar dx^{'k}$ should be considered as an implicit form of the transformation law of embedding vectors.

Let us find an explisit form of this law. Equation \eqref{27} can be rewritten as
\begin{equation*}
\bar dx^i=\left(\frac{\partial x^i}{\partial x^{'k}}+\hat b'_{\ k} x^i\right)\left(\bar dx^{'k}-dx^l\hat b_l x^{'k}\right)=
\end{equation*}
\begin{equation*}
=\frac{\partial x^i}{\partial x^{'k}}\bar dx^{'k}+\bar dx^{'k}\hat b'_{\ k}x^i-dx^l\frac{\bar \partial x^i}{\partial x^{'k}} \hat b_l x^{'k},
\end{equation*}
and opening the last terms, we get\footnote{All transformations in the calculations in this section should be performed {\em without using} the orthogonality $dx^l\hat b_l=0$, otherwise it is easy to lose the anisotropic contributions.}
\begin{equation*}
\bar dx^i=\frac{\partial x^i}{\partial x^{'k}}\ \bar dx^{'k}+dx^{'k}\hat b'_{\ k}x^i-dx^l\frac{\partial x^i}{\partial x^{'k}}\ \hat b_l x^{'k}.
\end{equation*}
Finally, noting that the matrix $\partial x^i/\partial x^{'k}=const(u^l)$, i. e. {\em commutes} with the direction operators $\partial/\partial u^m$ in $\hat b_l$, we have
\begin{equation*}
\bar dx^i=\frac{\partial x^i}{\partial x^{'k}}\ \bar dx^{'k}+dx^{'k}\left(\hat b'_{\ k}-\frac{\partial x^l}{\partial x^{'k}}\ \hat b_l\right) x^i.
\end{equation*}

From this we can already see that the last two terms compensate each other when choosing a solution in the form (i. e., the operator $\hat b_i$ is a vector)
\begin{equation}\label{28}
\bar dx^i=\frac{\partial x^i}{\partial x^{'k}}\ \bar dx^{'k},
\end{equation}
and Eq. \eqref{28} itself may be interpreted as the sought -- for {\em transformation law} of the infinitesimal vectors $\bar dx^i$ and $\bar dx^{'k}$. In this case, the operator $\hat b_l$ is a {\em vector} (as
it should be) \eqref{5}, and the anisotropic differentials $\bar dx^i$ are also transformed as isotropic ones, $dx^i$\footnote{The result obtained was generally expected: in the Eulerian representation, the anisotropy
vector $u^i$ is a function of coordinates, that is, the developed anisotropic geometry is Riemann, $g_{ik}(x,u)=g_{ik}[x,u(x)]$.}.

Thus, an arbitrary contravariant vector $A^i(x^k,\\u^l)$ should be understood as a directed quantity that transforms like full differentials \eqref{28} under the coordinate transformations \eqref{25}
\begin{equation}\label{29}
A^i=\frac{\partial x^i}{\partial x^{'k}}\ A^{'k}.
\end{equation}

The requirement that the scalar product of arbitrary vectors $A_iB^i=A'_kB'^k$ is invariable gives the transformation law of the covariant vector $A_i(x^k,u^l)$
\begin{equation}\label{30}
A_i=\frac{\partial x^{'k}}{\partial x^i}\ A'_k.
\end{equation}

It is now clear how \eqref{25} transforms the metric tensor and keep the scalar $ds^2$ unchanged, and also that $u^i$ is really a vector and is transformed according to \eqref{29} (just as in the Riemann case).

The anisotropic gradient of the embedding scalar $\varphi(x^i,u^k)=\varphi'(x^{'i},u^{'k})$ can also be defined  using the anisotropic differential
\begin{equation*}
\bar d\varphi=\frac{\bar \partial \varphi}{\partial x^i}\ dx^i\quad \Rightarrow\quad \frac{\bar \partial \varphi}{\partial x^i}=\frac{\partial x^{'k}}{\partial x^i}
\frac{\bar \partial \varphi}{\partial x^{'k}}\quad \Rightarrow
\end{equation*}
\begin{equation}\label{31}
  \quad \frac{\bar \partial}{\partial x^i}=\frac{\partial x^{'k}}{\partial x^i}\frac{\bar \partial}{\partial x^{'k}}
\end{equation}
and it is clear that the anisotropic gradient operator is a vector in the {\em Embedding}.

It is easy to show that the anisotropic differenti\-al of a vector and its anisotropic derivative are not a vector and tensor, respectively,
\begin{equation}\label{32}
\frac{\bar\partial A_i}{\partial x^k}=\frac{\partial x^{'l}}{\partial x^i}\frac{\partial x^{'m}}{\partial x^k}\frac{\bar \partial A'_{\ l}}{\partial x^{'m}}+
A'_l\frac{\partial^2 x^{'l}}{\partial x^k\partial x^i}
\end{equation}
due to the presence of terms with second-order derivatives\footnote{Here $\bar \partial\partial x^{'l}/\partial x^k\partial x^i=\partial^2 x^{'l}/\partial x^k\partial x^i$ since $\hat b_k\partial x^{'l}/\partial x^i=0$,
because $\partial x^{'l}/\partial x^i=const(u^i)$.}. However, adding the term $-\Gamma^l_{\ ik}A_l$  to the right and left-hand sides of \eqref{32} allows us to define a covariant anisotropic derivative of a vector and
its covariant anisotropic differential
\begin{equation*}
A_{i|k}\equiv\frac{\bar\partial A_i}{\partial x^k}-\Gamma^l_{\ ik}A_l, \quad A_{i|k}=\frac{\partial x^{'l}}{\partial x^i}\frac{\partial x^{'m}}{\partial x^k}\ A'_{l|m},
\end{equation*}
\begin{equation}\label{33}
\bar DA_i=A_{i|k}dx^k
\end{equation}
with the transformation law of the quantity $\Gamma^i_{\ kl}$
\begin{equation}\label{34}
\Gamma^i_{\ kl}=\frac{\partial x^i}{\partial x^{'m}}\left(\frac{\partial x^{'n}}{\partial x^k}\frac{\partial x^{'p}}{\partial x^l}\Gamma'^m_{\ np}+
\frac{\partial^2 x^{'m}}{\partial x^k\partial x^l}\right),
\end{equation}
which makes us identify the latter as the connection of the {\em Embedding} $M_e$.

\subsection{\textmd{Curvatures of {\em Embedding}}}
\label{Curv}
\quad It is natural to expect that the solution of the MES curvature problem is a certain generalization of the Riemann concept in the spirit of the develop\-ed ideas. The definition of the MES curvature tensor, in
contrast to the homogeneous Riemann case (a unique manifold), must take into account the structure and internal mobility of the {\em Embedd\-ing} $M_e$. It is a quite clear that the desired generali\-zation must
coincide with the Riemann concept in the homogeneous case and give nonzero value for the {\em Embedding} even for the flat spaces.

On the other hand, in the covariant differential \eqref{33} of an arbitrary $A_i(x,u)$
\begin{equation*}
 \bar DA_i=\left[\left(\partial/\partial x^k+\hat b_k\right)A_i-\Gamma^l_{\ ik}A_l\right]dx^k,
\end{equation*}
the second and third terms describe increments due to both the mobility and the structure of the {\em Embedding}
\begin{eqnarray}\label{35}
  \nonumber
  \bar DA_i=\left(\partial A_i/\partial x^k\right)dx^k-\delta A_i, \\
  \delta A_i=\left(\Gamma^l_{\ ik}A_l-\hat b_kA_i\right)dx^k.
\end{eqnarray}

Using these contributions, we con calculate the local circulation $\Delta A_i$ caused by the curvature
\begin{equation*}
 \Delta A_k=\oint_{C}\delta A_k=\oint_{C}\left(\Gamma^i_{\ kl}A_i-\hat b_lA_k\right)dx^l,
\end{equation*}
at the centre of a small closed contour $C$, see, e. g., \cite{LL1988}. (It should be noted here that the orthogonality of the operator $\hat b_l$ \eqref{5} of the transfer curve, $\hat b_lu^l=0$, requires care in further
transfor\-mations: one should not omit the corresponding terms until the result is obtained.)

Applying the Gauss theorem, we can write for the increments
\begin{equation*}
 \Delta A_k=(1/2)\left[\partial\left(\Gamma^i_{\ km}A_i-\hat b_m A_k\right)/\partial x^l-\right.
\end{equation*}
\begin{equation}\label{36}
 \left.-\partial\left(\Gamma^i_{\ kl}A_i-\hat b_l A_k\right)/\partial x^m\right]\Delta f^{lm},
\end{equation}
where, as usual, $\Delta f^{lm}\equiv \Delta x^l\Delta x^{'m}-\Delta x^m\Delta x^{'l}$ is the area inside the contour $C$.

The coordinate derivatives in this expression are determined by {\em only} the curvature of space, so \eqref{35} can be considered as a definition of the coordinate derivative of the vector $A_i(x,u)$ at any point
inside $C$:
\begin{equation*}
 \partial A_i/\partial x^k=\Gamma^l_{\ ik}A_l-\hat b_kA_i.
\end{equation*}

Using this relation, and taking into account the linearity of $\hat b_i$ and commutativity with $\partial/\partial x^i$, the expression \eqref{36} can be transformed to
\begin{equation}\label{37}
 \Delta A_k=(1/2)\left(\bar P^i_{\ klm}A_i-2\hat b_{[l}\hat b_{m]}A_k\right)\Delta f^{lm},
\end{equation}
where the {\em cuvature tensor} of the {\em Emdedding} is\footnote{$\Delta A^k=-(1/2)\left(\bar P^k_{\ ilm}A^i+2\hat b_{[l}\hat b_{m]}A^k\right)\Delta f^{lm}$.}
\begin{equation*}
 \bar P^i_{\ klm}=\frac{\bar\partial\Gamma^i_{\ km}}{\partial x^l}-\frac{\bar\partial\Gamma^i_{\ kl}}{\partial x^m}+2\Gamma^i_{\ n[l}\Gamma^n_{\ m]k},
\end{equation*}
\begin{equation}\label{38}
 \bar P_{iklm}=\frac{\bar\partial\Gamma_{i,km}}{\partial x^l}-\frac{\bar\partial\Gamma_{i,kl}}{\partial x^m}+2\Gamma^n_{\ m[i}\Gamma_{n,k]l}.
\end{equation}

Equality to zero of the local metric circulation determines the even (symmetric part) of the curva\-ture tensor with respect to the {\em first pair} of indices,
\begin{equation*}
 \Delta g_{ik}=\left(\bar P_{(ik)lm}-\hat b_{[l}\hat b_{m]}g_{ik}\right)\Delta f^{lm}=0\quad\Rightarrow\quad
\end{equation*}
\begin{equation}\label{39}
 \bar P_{(ik)lm}=\hat b_{[l}\hat b_{m]}g_{ik}\equiv df_{lm,ik}/ds,
\end{equation}
see \eqref{24}. As it easy to verify, \eqref{39} is an identity.

The second part of the curvature tensor, which is antisymmetric in the first pair of indices, is obviously the {\em Riemann tensor} of the {\em Embedding}
\begin{equation*}
 \bar R_{iklm}\equiv\bar P_{[ik]lm}\quad \Rightarrow
\end{equation*}
\begin{equation}\label{40}
 \bar R_{iklm}=\frac{\bar\partial\Gamma_{[i,k]m}}{\partial x^l}-\frac{\bar\partial\Gamma_{[i,k]l}}{\partial x^m}+2\Gamma^n_{\ m[i}\Gamma_{n,k]l}.
\end{equation}

The Riemann tensor splits into three terms $R_{iklm}$,  $r_{iklm}$ and $p_{iklm}$
\begin{equation}\label{41}
R_{iklm}=\frac{\partial \gamma_{i,km}}{\partial x^l}-\frac{\partial \gamma_{i,kl}}{\partial x^m}+2\gamma^n_{\ m[i}\gamma_{n,k]l},
\end{equation}
\begin{equation}\label{42}
r_{iklm}=2\left(\hat b_{[l}\omega_{[i,k]m]}+\omega^n_{\ m[i}\omega_{n,k]l}\right),
\end{equation}
correspond to the sets $M_s$ and $M_c$ respectively, while the {\em mixed} tensor, corresponding to the elect\-rogravity of the  {\em Embedding},
\begin{equation*}
  p_{iklm}\equiv 2\left(\frac{\partial \omega_{[i,k]m]}}{\partial x^{[l}}+\hat b_{[l}\gamma_{[i,k]m]}-\right.
\end{equation*}
\begin{equation}\label{42v}
  \qquad \qquad \qquad \qquad \qquad \left. -2\gamma^n_{\  \ [l[i}\omega_{n,k]m]}\right),
\end{equation}
\begin{equation}\label{42vv}
  p_{iklm}=-p_{kilm}=-p_{ikml},\quad p_{iklm}=p_{lmik},
\end{equation}
by definition of the operators \eqref{5n}-\eqref{5nn}, is {\em identically equal to zero}.  (The MES {\em has no} intersections of $M_s$ and $M_c$.)

The symmetry under permutation of the index pair $\{ik\}\leftrightarrows\{lm\}$ follows from commutativity of the operators $\partial/\partial x^i$ and $\hat b_k$, in particular,
\begin{eqnarray}
\nonumber
  \hat b_{[i}\gamma_{[l,m]k]}=\partial \omega_{[i,k]m]}/\partial x^{[l},\\
  \label{42vvv}
  \partial \omega_{[l,m]k]}/\partial x^{[i}=\hat b_{[l}\gamma_{[i,k]m]} .
\end{eqnarray}

It is quite clear that the contractions of the curvature tensor $M_s$ are the Ricci tensor and scalar, defined in the standard way
\begin{equation*}
   R_{ik}=g^{lm}R_{limk},\quad R_{ik}=R_{ki},\quad R=g^{ik}R_{ik},
\end{equation*}
\begin{equation}\label{41v}
  R_{ik}=\frac{\partial \gamma^l_{\ ik}}{\partial x^l}-\frac{\partial \gamma^l_{\ il}}{\partial x^k}+\gamma^l_{\ ik}\gamma^m_{\ lm}-\gamma^l_{\ im}\gamma^m_{\ kl},
\end{equation}
they also have standard properties.

Using \eqref{19} and \eqref{24}, it is easy to verify that $r_{iklm}$ does not have the characteristic index sym\-metry of the Riemann tensor with respect to permutation of the index pair
$\{ik\}\leftrightarrows\{lm\}$ (the first term in \eqref{42}):
\begin{equation*}
r_{iklm}-r_{lmik}=\hat b_{[l}\hat b_{k]}g_{im}+\hat b_{[i}\hat b_{l]}g_{km}+\hat b_{[k}\hat b_{m]}g_{il}+
\end{equation*}
\begin{equation*}
  +\hat b_{[m}\hat b_{i]}g_{kl}=4df_{[i[l,m]k]}/ds,\quad f_{[i[l,m]k]}\equiv-f_{[l[i,k]m]},
\end{equation*}
where antysymmetrization with respect to the inner and outer brackets is performed separately: $\left[.[.,.].\right]$.

This shows that the asymmetry is caused by the non-commutativity of the operators $\hat b_i$ \eqref{24}. It is also clear that the symmetry of the $r_{iklm}$ under discussion can be provided by
the general requirement
\begin{equation*}
  df_{[l[i,k]m]}/ds=0 \quad \Rightarrow
\end{equation*}
\begin{equation}\label{43}
 \quad f_{[l[i,k]m]}=C_{iklm}=const(x),
\end{equation}
where $s$ is the interval of parallel transfer along an infinitesimal contour around any point, and $C_{iklm}$ is a certain {\em structural} constant of the {\em Embedding} geometry. The validity of \eqref{43}
provides a Riemann version of index symmetry.

The absence of the $\{ik\}\leftrightarrows\{lm\}$ symmetry also results in {\em violation } of the Ricci identity. Let $..._{\ldots \ ikl\circlearrowleft \ \ldots}$ denote a cyclic sum over the indices $i,k,l$, then
\begin{equation*}
  r_{iklm\circlearrowleft}\equiv r_{iklm}+r_{imkl}+r_{ilmk}=\hat b_k\omega_{[l,m]i}+
\end{equation*}
\begin{equation}\label{44}
+\hat b_m\omega_{[k,l]i}+\hat b_l\omega_{[m,k]i}=-df_{kl,m\circlearrowleft i}/ds.
\end{equation}
The identity is restored if the condition \eqref{43} is met, since
\begin{equation*}
 4f_{[k[l,m]i]}\equiv f_{kl,m\circlearrowleft i}-f_{il,m\circlearrowleft k}.
\end{equation*}

Note that due to the non-commutativity \eqref{24}, an analog of the Bianchi identity for $r_{iklm}$ {\em does not hold}:
\begin{equation*}
  r_{nikl:m\circlearrowleft}\equiv r_{nikl:m}+ r_{nimk:l}+ r_{nilm:k}=
\end{equation*}
\begin{equation*}
\quad \quad=2\left(\hat b_{[i}\dot{f}_{kl,m\circlearrowleft n]}-\omega^p_{\ m[i}\dot{f}_{kl,pn]}-\right.
\end{equation*}
\begin{equation*}
\quad \quad \quad \quad \left.-\omega^p_{\ l[i}\dot{f}_{mk,pn]}-\omega^p_{\ k[i}\dot{f}_{lm,pn]}\right),
\end{equation*}
where the overdot means $d/ds$.

The lack of symmetry under permutation of pairs in $r_{iklm}$ also lead to an {\em asymmetry} of $r_{ik}\equiv g^{lm}r_{limk}$
\begin{equation*}
 2r_{[ik]}=g^{lm}\left[-\hat b_m\omega_{[i,k]l}+\hat b_{[i}\left(\omega_{l,mk]}-\omega_{k],lm}\right)\right],
\end{equation*}
\begin{equation*}
 2r_{(ik)}=g^{lm}\left[\hat b_m\left(\omega_{l,ik}-\omega_{(i,k)l}\right)+\hat b_{(i}\left(\omega_{k),lm}-\right.\right.
\end{equation*}
\begin{equation}\label{45}
 \left.\left.-\omega_{l,mk)}\right)-2\left(\omega^n_{\ \ ik}\omega_{n,lm}-\omega^n_{\ \ il}\omega_{n,km}\right)\right].
\end{equation}

Finally, let us present the explicit form of the curvature of $M_c$
\begin{equation*}
  r=g^{ik}r_{ik}=2\hat b_i\omega^{[i,k]}_{\ \ \ \ k}+\omega_{i,kl}\omega^{l,ki}-\omega^i_{\ ik}\omega_{l,}^{\ \ lk}\equiv
\end{equation*}
\begin{equation}\label{46}
 \equiv 4u^i\;\partial f_{i\ \ \ \ l}^{\;[k,l]}/\partial x^k-\xi_{ik}u^iu^k,
\end{equation}
where ($\xi\equiv\xi^i_{\ i}$)
\begin{eqnarray}\label{47}
  \nonumber
  \xi_{ik}=f_{il,m}^{\ \ \ m}f_{k\ \ \ n}^{\ l,n}+f_{il,mn}\left(f_k^{\ l,mn}-2f_k^{\ m,ln}\right), \\
  \xi=f_{lm,n}^{\ \ \ \ n}f^{lm,p}_{\ \ \ \ \ p}+f_{lm,np}\left(f^{lm,np}-2f^{ln,mp}\right).
\end{eqnarray}

Despite the identical vanishing of the {\em mixed} curvature tensor \eqref{42v}, we give an explicit form of the remaining curvatures as well:
\begin{equation*}
  p_{ik}=g^{lm} p_{limk},\quad  p_{ik}\neq p_{ki},\quad p=g^{ik}p_{ik},
 \end{equation*}
 \begin{equation*}
  p_{ik}=2\left(\partial \omega^{[l}_{\ \ i]k]}/\partial x^{[l} +\hat b_{[l}\gamma^{[l}_{\ \ i]k]} +\omega_{[i,l][m}\gamma^{m,l}_{\ \ \ \ k]}+\right.
 \end{equation*}
 \begin{equation*}
   \left.+\gamma_{[i,l][m}\omega^{m,l}_{\ \ \ \ k]}+\omega_{(i,l)[m}\gamma^{l,m}_{\ \ \ \ k]}+\gamma_{(i,l)[m}\omega^{l,m}_{\ \ \ \ k]}\right),
 \end{equation*}
\begin{equation*}
  p=2\left(\partial \omega^{[i,k]}_{\ \ \ \  k]}/\partial x^{[i}+\hat b_{[i}\gamma^{[i,k]}_{\ \ \ \ k]} +\right.
 \end{equation*}
\begin{equation}\label{43v}
  \left. \qquad\qquad\qquad+\omega_{i,kl}\gamma^{(k,l)i}-\omega^i_{\ ik}\gamma_l,^{lk}\right)
\end{equation}
where antisymmetrization by inner and outer brackets is performed separately, as usual: $\left[.[.,.].\right]$.

It should also be mentioned that since the division of algebraic objects of the geometry  into tensor and non-tensor quantities is determined by their transformation laws \eqref{28} -- \eqref{34} with \eqref{25},
the operator $\partial/\partial u^i$ must be understood as a {\em vector} due to the vector nature $u^i$ (similarly to $\partial/\partial x^i$),
\begin{equation*}
\partial/\partial u^i=\left(\partial x'^k/\partial x^i\right)\partial/\partial u'^k.
\end{equation*}
And from here it is already easy to prove that $\hat b_{ik}$, $\hat b_i$, $\bar \partial/\partial x^i$, the generally covariant derivatives of vectors like $A_{i|[k|l]}$, the quantity $p_{iklm}$, etc., are the tensors of
different ranks. Unlike that, $f_{ik,lm}$ is not tensor:
\begin{equation*}
  f_{ik,lm}=\frac{\partial x'^n}{\partial x^i}\frac{\partial x'^p}{\partial x^k}\frac{\partial x'^r}{\partial x^l}\frac{\partial x'^s}{\partial x^m}f'_{np,rs}-
\end{equation*}
\begin{equation}\label{48}
-2c'_{n,pr}\frac{\partial x'^n}{\partial x^{[i}}\frac{\partial}{\partial x^{k]}}\left(\frac{\partial x'^p}{\partial x^l}\frac{\partial x'^r}{\partial x^m}\right).
\end{equation}

\section{\textmd{FIELD EQUATIONS}}
\label{FieldEq}
\quad One of the main goals of this paper is to derive the field equations of the MES model. To solve the problem, we use the Lagrangian description of a classical physical system and the least action principle. In this
case, the system is compared with a matter-field Lagrangian density of GR type, consisting of additive contributions of free matter and fields. Moreover, in this case, the mass density in $\Lambda_0$ should be understood
as a  {\em seed} ("bare"\ ) mass $\mu_0$, without any contribution of the {\em proper} fields: otherwise, in the field components of the Lagrangian, these fields will be included {\em twice}\footnote{Unlike GR, here {\em both} classical fields are geometrized.}.

On the other hand, it is known that a massless electric charge does not exist in nature. That is, following classical electrodynamics, it is necessary to assume that every charge distribution is simul\-taneously a certain mass distribution. Therefore, assuming the existence of $\mu_0$, we automatically assume the existence of $\rho_0$.

Lastly, there is the world constant $\pm\sqrt{k}$, which is the same in dimension and sign as $\beta=\rho/\mu$, \eqref{2}.

In summary, we can accept the hypothesis of {\em seed charged} matter: {\em such matter does exist and also moves along geodesics of the Embedding $M_e$}, \eqref{4}, \eqref{16}, \eqref{17} and it
holds
\begin{equation}\label{49}
\rho_0/\mu_0=\pm\sqrt{k},
\end{equation}
where $\rho_0$, $\mu_0$ are the distribution densities of the seed charge and mass. Obviously, $\pm\sqrt{k}$ must also be understood as a {\em charge-mass characteristic} of a physical vacuum of the
{\em Embedding}\footnote{The experimental determination of the sign of $\pm\sqrt{k}$ is important: it must be determined by the baryon asymmetry of the Universe.}.

However, in the initial representations of the metric \eqref{2}, the "material"\ analogs of the potentials $\vartheta_i$ and $c_{ikl}$, \eqref{3}, the magnitude of $f_{ik,lm}$, \eqref{19}, and the scalar $\beta$,
the quantities $\rho$ and $\mu$ are understood to be {\em real} densities. That is, both the seed $\rho_0/\mu_0$ \eqref{49}, and the real $\rho/\mu$ matter must move along {\em coinciding} geodesics!

In this case, the identification of $\rho/\mu$ and $\pm\sqrt{k}$ is completely impossible: it would mean replacing the spectrum $dq/dm=q/m$ \cite{LL1988} of real charge-mass distributions with only two(!) {\em fundamental} values. In addition, even if we assume that $\beta=\rho/\mu$ of real matter is somehow configured from the constants $\pm\sqrt{k}$, there is still a very serious objection caused by the huge specific
charges of elementary particles. For example, for electron $(|q_e|/m_e)/\sqrt{k}\simeq 2\cdot 10^{21}$, which means that its real acceleration in an electromagnetic field should exceed the acceleration of its own
seed matter by 21 orders of magnitude.

This glaring contradiction is removed by the {\em field hypothesis:} the real mass is the seed mass with proper field contributions,
\begin{equation}\label{49v}
  \mu\simeq\mu_0+w_{(pf)}/c^2,
\end{equation}
where $w_{(pf)}$ is the energy density  of the {\em proper} fields of matter, the electromagnetic and gravity ones. As we know, these field contributions are of different signs... That is, the field hypothesis
suggests that the action of an {\em external} field $A_i$ on real matter $\beta Au$ \eqref{2}, can be understood as the action of the {\em complete} field with a potentional $a_i$ (the proper "plus"\ external
ones\footnote{The field superposition only works for weak fields.}) on the {\em seed} matter,
\begin{equation}\label{50}
  (\rho/\mu)Au\equiv \pm\sqrt{k}\, au.
\end{equation}

{\em Thus, taking into account the self-action allows us to treat the trajectory of real matter with $\rho/\mu$ in the external field $A_i$ as a trajectory of seed matter, $\pm\sqrt{k}$, in the total field $a_i$.}

According to GR, the curvatures $R$ and $r$ must be used for the field Lagrangian densities, and the gravitation term must be included with the weight factor $c^4/16\pi k$. Accordingly, for $r$, we should choose the
factor $c^4/16\pi \beta^2$ because the analog of the Maxwell fild tensor $F_{ik}$ in the {\em Embedding} is $c^2f_{ik,lm}/\beta$.

Some difficulty in interpreting the value of $c^4r/16\pi \beta^2$ as a Lagrangian electromagnetic field density arises due to an unusual explicit form of curvature $r$ \eqref{46}: it is determined not by the quantity
$f_{ik,lm}$ itself, but by its {\em projection} on the velocity vector $u^i$. However, for the structure  of the MES space adopted in this paper, the {\em Embed\- ding} of the sets $M_s$ and $M_c$, with $M_c$ {\em
moving with the velocity $u^i$}, this fact is both physically acceptable and necessary.

Finally, the geometrization requires emphasi\-zing the {\em geometric content} of the Lagrangian formalism. This requirement can be taken into account by noticing  that the action of the physical system under discussion
clearly depends on the local "density"\ of points and the local anisotropy of the {\em Embedding} $M_e$. The simplest quantities describing these dependences are: $g_{ik}$, $c_{i,kl}$ and $u^i$.

\subsection{\textmd{The System Action}}
\label{SysAct}
\quad The equivalence of Lagrangian and Eulerian description of a continuous medium allows us to use the variational principle of least action for the matter-field system under consideration to obtain the field equations.
The system action in the {\em Embedding} $M_e$ must obviously contain the Lagrangian density of free {\em seed} (isotropic) matter and the spatial curvatures $R$ \eqref{41v} (gravitational) and $r$
\eqref{46} - \eqref{47} (electric) as the field densities. Moreover, the signs of the field terms should be opposite, as should be in the field theory.

As independent variation parameters, one sho\-uld choose the metric (for a fixed direction),  $g^{ik}(x,u=const)$, and this direction itself at a fixed point of the {\em Embedding} -- the local velocity of matter $u^i(x)$, \cite{Nskv2007}. Since these parameters are normalized, $g_{ik}g^{ik}-4\equiv 0$ and $u_iu^i-1\equiv 0$, these norms can be considered as holonomic relations imposed on the  functional. Then the search for the action extremum reduces to the problem of finding conditional functional extremum, which, as is well known, is solved by the method of indefinite Lagrange multipliers \cite{Smirnov1956}.

For convenience, as will become clear later, it is reasonable to immediately modify the system Lagrangian by explicitly taking these relations into account, which is obviously equivalent to a similar modification in the Lagrange equations.  Thus, the "working"\ form of the desired action looks as
\begin{equation*}
 cS=\int\left[\Lambda_0-\frac{c^4}{16\pi}\left(\frac{R}{k}-\frac{r}{\beta^2}\right)+\lambda_g(g_{ik}g^{ik}-4)+\right.
\end{equation*}
\begin{equation}\label{51}
 \qquad \left.+\lambda_u(u_iu^i-1)\right]\sqrt{-g}d\Omega,
\end{equation}
where the Lagrangian density of free isotopic {\em seed} matter is\footnote{$\Lambda_0$ is a generally covariant scalar: $\Lambda_0/c^2=-(dm_0/\sqrt{\gamma}dV)\cdot (ds/\sqrt{g_{00}}dx^0)=-ds\, dm_0/\sqrt{-g}\,d\Omega,$ where $ds$, $dm_0$ и $\sqrt{-g}\, d\Omega$ are scalars \cite{Vlad2009}.}
\begin{equation}\label{52}
\Lambda_0=-\mu_0c^2ds/\sqrt{g_{00}}\;dx^0,
\end{equation}
\begin{equation}\label{53}
\lambda_g\neq \lambda_g(g^{ik}),\; \lambda_u\neq\lambda_u(u^i),
\end{equation}
are Lagrangian multipliers of the variation task,
\begin{equation*}
g=\det[g_{ik}], \, d\Omega=dx^0dx^1dx^2dx^3,
\end{equation*}
\begin{equation}\label{54}
  \varkappa=8\pi k/c^4,\, \beta=\rho/\mu,
\end{equation}
$g$ -- is the metric determinant, and $d\Omega$ is the 4-volume element.

As mentioned above, the {\em independent} variation in $g^{ik}$ and $u^i$ has the necessary geometrical mean\-ing required by the geometrization problem: in terms of two-dimensional density of points
{\em Embed\-ding} is isotropic, and in the direction it is anisotro\-pic. The conditions of the corresponding extremum read
\begin{eqnarray}\label{55}
  \nonumber
  \delta S[g^{ik},u^l]_{|u^l=const}=0, \\
  \delta S[g^{ik},u^l]_{|g^{ik}=const}=0.
\end{eqnarray}

One can guess that the extrema \eqref{55} give Einstein-type and Maxwell-type equations. Indeed, varying the functional \eqref{51} in $g^{ik}(x^l,u^m=const)$ ($c_{i,kl}=0$ -- no electricity) leads to equation
of gravity for an {\em arbitrary} chosen direction $u^m(x)=const$, i. e., to {\em anisotropic} gravity. Conversely, variation of \eqref{51} in an {\em explicit} dependence on $u^m$ for {\em arbitrary} selected
$g^{ik}(x^l,u^m)=const$  leads to the vector equation of electricity under arbitrary gravity. Obviously, the equations obtained in this way (the tensor and vector ones) must be solved simultaneously.

{\em Remark.} The density of points of the {\em Embedding} and the direction in any coordinate system point $x$ can also be the pair of variables $(g_{ik},u_l)$, or equally $(g^{ik},u^l)$: the choice is completely
conditio\-nal. However, using the pair $(g^{ik},u^l)$ is preferable, by analogy with the choice of $g^{ik}$ in the derivation of Einstein's equation (the Hilbert version).

The version of mixed pair, for example, $(g^{ik},u_i=g_{ik}u^k)$ is unfortunate because variation in $u_i$ {\em will not be independent}, it is based on the product "density of points $\otimes$ direction"\ .
The result will be opaque and unclear! Thus, for selected variable $g^{ik}$, independent anisotropy variable should be $u^i$ rather than $u_i$.

\subsection{\textmd{The Electricity Equations}}
\label{ElmEq}
\quad As will be seen later, it is convenient to start the search for field equations with electromagnetism, \eqref{55}:
$$
\delta S[g^{ik},u^l]_{|g^{ik}=const}=0
$$
 the variation of the action of the system \eqref{51} in the direction (by an explicit dependence on $u^i$) must be zero.

To begin with, we briefly describe the method of indefinite Lagrangian multipliers \cite{Smirnov1956} applied to variational problem of the type of \eqref{55}. Since $g^{ik}$ and $u^l$ are chosen as independent variational variables, the variations of the covariant compon\-ents $g_{ik}$ and $u_l$, in accordance with the method, become indefinite:modificafication of the original functional to get the working one, \eqref{51}, {\em "removes"\
the normalizations}. And a convenient choice of these variations, for example, $\delta g_{ik}=0$ and $\delta u_l=0$ ($g_{ik}$ and $u_l$ are assumed to be {\em constants} of variations), transfers the resulting uncertainties
to the corresponding Lagrangian multipliers, $\lambda_g$ and $\lambda_u$, Eq. \eqref{53}. The latter are selected based on any task-specific considerations.

Further, in the results of variation, the norms of variation variables are restored. (Exam\-ple: the above-mentioned Hilbert's derivation of the Eins\-tein equations , variation in $g^{ik}$, is realized at $\delta g_{ik}=0$,
the choice of the Lagrangian multiplier is a cosmological "constant"\ , and in the resulting equations we have again $g_{ik}g^{ik}=4$.)

So, we vary in the direction (in $u^i$) the first term in \eqref{51}, assuming that $g^{ik}=const$ is the variation constant, $\delta g^{ik}/\delta u^l=0$. To do that, we write $\Lambda_0$ from \eqref{52}, $ds=+\sqrt{dx_idx^i}$, in a convenient form
\begin{equation*}
  \Lambda_0=-\mu_0c^2\sqrt{u_iu^i}/\sqrt{g_{00}}\ u^0.
\end{equation*}
The method requires variation with a {\em non-norma\-lized} scalar $u_ku^k$ and $u_l=const$ (the covariant velocity is a constant of variation, $\delta u_l/\delta u^i=0$). It is not hard to see that the denominator $\Lambda_0$
is a constant of variation since $\sqrt{g_{00}}\, u^0=\sqrt{g^{00}}\, u_0$, for example, at $u^\alpha=0$. Therefore,
\begin{equation*}
  \delta \Lambda_0=\Lambda_0\delta \ln{\sqrt{u_ku^k}}\, \Rightarrow\,  \delta \Lambda_0/\delta u^i=\Lambda_0u_i/2.
\end{equation*}
Then variation of the functional \eqref{51} in an {\em explicit} dependence on the velocity \eqref{55} (also $\delta\beta/\delta u^i=0$) gives
\begin{equation*}
  \delta\Lambda_0/\delta u^i+(c^4/16\pi\beta^2)\delta r/\delta u^i+\lambda_uu_i=0 \quad \Rightarrow
\end{equation*}
\begin{equation}\label{56}
 \quad \delta r/\delta u^i=-8\pi\beta^2c^{-4}(\Lambda_0+2\lambda_u)u_i.
\end{equation}

Having found $\delta r/\delta u^i$ from \eqref{46}, we obtain
\begin{equation*}
2\partial f_{i\ \ \ \ \ l}^{\ [k,l]}/\partial x^k-\xi_{ik}u^k=-4\pi\beta^2c^{-4}\left(\Lambda_0+2\lambda_u\right)u_i.
\end{equation*}
or
\begin{equation}\label{57}
2\,\frac{\partial f_{i\ \ \ \ \ l}^{\ [k,l]}}{\partial x^k}=-4\pi\beta^2c^{-4}\left(\Lambda_0g_i^{\ k}+a_i^{\ k}\right)u_k,
\end{equation}
where
\begin{equation}\label{58}
a_{ik}\equiv\left(-\xi_{ik}+8\pi\beta^2c^{-4} g_{ik}\lambda_u\right)/4\pi\beta^2c^{-4}.
\end{equation}

From the definition \eqref{58}, it can be seen that choosing the constant $\lambda_u$ as the contraction
\begin{equation}\label{59}
\lambda_u\sim \xi/32\pi\beta^2c^{-4}
\end{equation}
makes the quantity $a_{ik}$ similar to the Maxwellian energy-momentum tensor of the electromagnetic field. However, as will be clear a bit later, this Lagrangian constant is also related to the gravita\-tional curvature $R$
{\em Embedding}. To clarify its choice, one should derive the equations of gravity.

{\bf Remark.}

Strictly speaking, Eqs. \eqref{57} are not analogous to Maxwell's field equations for two reasons. First, the "potential"\ $c_{i,kl}=\partial g_{kl}/\partial u^i$ implicitly depends on the matter characteristic $\beta=\rho/\mu$ (see, e. g., $c_{i,00}$ in the case of \eqref{2}), where $\mu$, in accordance with the field hypothesis, has a field contribution. The latter obviously creates a significant field nonlinearity. Secondly, the derivation method of
\eqref{57} is completely not canonical, although close to the latter: the vector $u$ is used, being a natural geometric indicator of anisotropy of the {\em Embedding} (the presence of the vector $A$). The choice of such a
(geometric) variation method is caused by the above-said, as well as by its simplicity and clarity. Since $u^i$ is a normalized vector (it has only three independent components), one of the four equations \eqref{57} is a
constraint.

However, an important task of the future "over-structure"\ over the geometric-physical "base"\ of this work is still to obtain a matter-field analog of Maxwell' equations in a canonical way, by varying the action of the system in $c_{i,kl}$. And not only because of the physical "validity"\ of the result, but also because of the latter's connection with the already discussed gauge properties of the {\em Embedding} anisotropy, see the end of Subsection \ref{CovDer}.

\subsection{\textmd{The Equations of Gravity}}
\label{GravEq}
\quad These equations are obtained as a result of the variation \eqref{55} of the action \eqref{51}
 $$\delta S[g^{ik},u^l]_{|u^l=const}=0,$$
moreover, the varied quantity $g^{ik}$ itself is a func\-tion of $u^l$ (the local direction). Therefore, (а) the velocity $u^l$ in all explicit and implicit occurrences in $S$ should be considered as an invariable constant,
$\delta u^i/\delta g^{kl}=0$ and (б) in the variable $g^{kl}$ and in the action $S$, the velocity $u^i$ must take admissible values, corresponding to the extremum \eqref{55}
$$\delta S[g^{ik},u^l]_{|g^{ik}=const}=0.$$

A natural way to take into account the require\-ment (b) is to use the already found "equation of motion of the system"\ \eqref{56} (the bottom line). This can be done in two ways: either {\em before} or {\em after} the
variation procedure. Due to linearity of the variation, these possibilities are {\em equivalent}, but the former is methodologically preferable. It allows one to exclude higher derivatives $\partial f_{i\ \ \ \ l}^{\;[k,l]}/\partial x^k$ \eqref{46} from the action $S$ and to write its field part in a canonical form -- only through {\em quadratic} expressions in $f_{ik,lm}$, which significantly simplifies the search for $\lambda_u$\footnote{Note that it would be error to understand \eqref{56} as a holonomic relation imposed on the variational variable $g^{ik}$: \eqref{56} only fixes the constant in  $g^{ik}(x^l,u^m=const)$.}.

So, we first reduce the action $S$ to a form quadratic in $f_{ik,lm}$. To do that, notice that by virtue of $r \sim u$, it always holds
\begin{equation*}
u^i(\delta r/\delta u^i)=r-\xi_{ik}u^iu^k,
\end{equation*}
and bycomparing it with \eqref{56}, we can express the curvature $r$ through $\xi_{ik}, \Lambda_0, \lambda_u$ and the local velocities:
\begin{equation}\label{60}
r=\xi_{ik}u^iu^k-8\pi\beta^2c^{-4}(\Lambda_0+2\lambda_u)u^2.
\end{equation}

Substituting this result in action \eqref{51} gives an expression quadratic in $f_{ik,lm}$
\begin{equation*}
  2cS=\int\left[\left(2-u^2\right)\Lambda_0-\frac{c^4}{8\pi}\left(\frac{R}{k}-\frac{\xi_{ik}u^iu^k}{\beta^2}\right)-\right.
\end{equation*}
\begin{equation*}
  \left.- 2\lambda_u+2\lambda_g\left(g_{ik}g^{ik}-4\right)\right]\sqrt{-g}d\Omega.
\end{equation*}

Hence, it is clear that a correct relationship between matter and gravity in the Lagrangian density of the system can only be achieved by choosing $-2\lambda_u\sim R/2\varkappa$. Therefore, the constant
$\lambda_u$ \eqref{59} is determined as
\begin{equation}\label{59v}
  \lambda_u=\frac{c^4}{32\pi}\left(\frac{\xi}{\beta^2}-\frac{R}{k}\right),
\end{equation}
which leads $a_{ik}$ \eqref{58} to the form
\begin{equation}\label{61}
a_{ik}=\frac{c^4}{4\pi\beta^2}\left(-\xi_{ik}+\frac{g_{ik}}{4}\xi\right)-g_{ik}\frac{R}{2\varkappa},\, a=-\frac{2R}{\varkappa},
\end{equation}
and the action \eqref{51} takes the desired form:
\begin{equation*}
  2cS=\int\left[\left(2-u^2\right)\Lambda_0-\frac{c^4}{16\pi}\left(\frac{R}{k}-\right.\right.
\end{equation*}
\begin{equation}\label{62}
  \left.\left.-\frac{2\xi_{ik}u^iu^k-\xi}{\beta^2}\right)+2\lambda_g\left(g_{ik}g^{ik}-4\right)\right]\sqrt{-g}d\Omega.
\end{equation}

A standard variation of \eqref{62} in $g^{ik}$ gives the Einstein-type equations
\begin{gather}
  \nonumber
   \qquad R_{ik}-\frac{g_{ik}}{2}R+\Lambda_c g_{ik}= \\
  \label{63}
    =\varkappa\left(t^{(0)}_{\ \ \ ik}+t^{(oem)}_{\ \ \ \ ik}+t^{(fem)}_{\ \ \ \ ik}\right),
\end{gather}
in which the energy-momentum tensors (EMTs) of the right-hand side are defined as \cite{LL1988}
\begin{equation*}
\frac12\sqrt{-g}\ t_{ik}=\frac{\partial (\sqrt{-g}\ \Lambda)}{\partial g^{ik}}-\frac{\partial}{\partial x^l}\frac{\partial(\sqrt{-g}\ \Lambda)}{\partial(\partial g^{ik}/\partial x^l)},
\end{equation*}
and $\Lambda_c=-4\varkappa\lambda_g$ is the {\em cosmological} "constant"\ .

Since $\partial\left(2-u^2\right)/\partial g^{ik}\equiv0$, because $u^2=g_{ik}u^iu^k$ we have
\begin{gather}
\label{64}
  t^{(0)}_{\ \ \ ik}= -u_iu_k\Lambda_0,\quad  t^{(0)}=-\Lambda_0
\end{gather}
an EMT of {\em seed} matter.

The EMT of the {\em proper} electromagnetic field of seed matter (contraction with the motion velocities) is\footnote{Here $\partial(u^lu^m)/\partial g^{ik}=0$ because $u^i$ is a constant of variation in \eqref{55}.}
\begin{equation*}
  8\pi\beta^2c^{-4} t^{(oem)}_{\ \ \ \ ik}=u^lu^m\left(2\partial/\partial g^{ik}-g_{ik}\right)\xi_{lm}\quad \Rightarrow
\end{equation*}
\begin{equation*}
  2\pi\beta^2c^{-4} t^{(oem)}_{\ \ \ \ ik}=u^lu^m\left[f_{ln,ik}f_{m \ \ \ \ p}^{\ \ n,p}+f_{li,n}^{\ \ \ (n}\cdot \right.
\end{equation*}
\begin{equation*}
  \left.\cdot f_{mk,p}^{\ \ \ \ \ p)}+2f_{ln,pi}\left(f_{m\ \ \ \ \ k}^{\ \ [n,p]}-f_{mk,}^{\ \ \ \ np}\right)-\frac{g_{ik}}{4}\;\xi_{lm}\right],
\end{equation*}
\begin{equation}\label{65}
4\pi\beta^2c^{-4} t^{(oem)}=\xi_{ik}u^iu^k.
\end{equation}

And finally, the EMT of the {\em free} electromagne\-tic field is
\begin{equation*}
16\pi\beta^2c^{-4} t^{(fem)}_{\ \ \ \ ik}=-\left(2\partial/\partial g^{ik}-g_{ik}\right)\xi\quad \Rightarrow
\end{equation*}
\begin{equation*}
4\pi\beta^2c^{-4} t^{(fem)}_{\ \ \ \ ik}=-\xi_{ik}+\frac{g_{ik}}{4}\xi-f_{lm,ik}f^{lm,n}_{\ \ \ \ \ n}-
\end{equation*}
\begin{equation*}
  \quad -2f_{lm,ni}f^{l[m,n]}_{\ \ \ \ \ \ k}-f_{il,mn}\left(f_k^{\ m,ln}+2f^{lm,n}_{\ \ \ \ \ k}\right),
\end{equation*}
\begin{equation}\label{66}
4\pi\beta^2c^{-4} t^{(fem)}=-\xi.
\end{equation}

Thus the sum of EMTs of the seed matter and its own electromagnetic field $t^{(0)}_{\ \ \ ik}+t^{(oem)}_{\ \ \ \ ik}$ is the EMT of "electromagnetically dressed"\ matter of the system. It corresponds to the "material"\
(non-field) terms the r.h.s. of the GR equations
\begin{equation}\label{67}
 T^{(mat)}_{\ \ \ \ \ ik}\equiv t^{(0)}_{\ \ \ ik}+t^{(oem)}_{\ \ \ \ \ ik}.
\end{equation}

\subsection{\textmd{Maxwell-Type Equation}}
\label{MaxvEq}
\quad To find the final form of a Maxwell-Type equati\-on, we substitute into \eqref{57} the expression for $a_{ik}$ \eqref{61}, in which $R/2\varkappa$ is found from the Einstein-type equations \eqref{63}:
\begin{gather*}
  R/2\varkappa=\frac{\Lambda_0}{2}+\frac{2\Lambda_c}{\varkappa}+\frac{c^4}{8\pi\beta^2}\left(\xi-\xi_{ik}u^iu^k\right),\\
  a_{ik}=-\left(\Lambda_0/2+2\Lambda_c/\varkappa\right)g_{ik}+\tau_{ik},
\end{gather*}
where
\begin{eqnarray}
\nonumber
    4\pi\beta^2c^{-4}\tau_{ik}&=&-\xi_{ik}+(g_{ik}/4)\left(2\xi_{lm}u^lu^m-\xi\right),\\
\label{68}
     2\pi\beta^2c^{-4}\tau&=&\xi_{ik}u^iu^k-\xi.
\end{eqnarray}

Then the desired equation reads
\begin{equation}\label{69}
  2\frac{\partial f_{i\ \ \ \ \ l}^{\ [k,l]}}{\partial x^k}=-\frac{4\pi\beta^2}{c^4}\left[\left(\frac{\Lambda_0}{2}-\frac{2\Lambda_c}{\varkappa}\right)g_i^{\ k}+\tau_i^{\ k}\right]u_k,
\end{equation}
or
\begin{equation}\label{70}
  2\frac{\partial f_{i\ \ \ \ \ l}^{\ [k,l]}}{\partial x^k}=\frac{4\pi\beta}{c^3\Lambda}\left[\left(\frac{\Lambda_0}{2}-\frac{2\Lambda_c}{\varkappa}\right)g_i^{\ k}+\tau_i^{\ k}\right]j_k,
\end{equation}
where $\Lambda$ is the Lagrangian density of free matter, and $j_i$ is the electric current density
\begin{equation}\label{71}
\Lambda = -\mu c^2\,ds/\sqrt{g_{00}}\,dx^0, \quad j_i=\rho c\,dx_i/\sqrt{g_{00}}\,dx^0.
\end{equation}

As a whole, the form of the vector equation found indicates its connection with Maxwell's equation and the {\em Embedding} anisotropy: the form of the r.h.s. indicates the field hypothesis of matter \eqref{49v}.
To confirm this, we consider a specific example of the {\em Embedding}.  A good example for that is a conform-Riemann metric.

\subsubsection{\textmd{A conformal-Riemann test}}
\label{test}
quad Consider the conformal-Riemann version  of the {\em Embedding} metric: in addition to its simplicity, it is of interest as a generalization of H. Weil's
idea of the conformal invariance of gravity \cite{Weil1918}. Let us assume that the metric has the form
\begin{equation}\label{71t}
  g_{ik}=\chi(x,u)g^{(r)}_{\ \ ik}(x),\quad \chi=(1+c^{-2}\beta\, Au)^2,
\end{equation}
where $g^{(r)}_{\ \ ik}(x)$ is the Riemann metric in $M_s$, and $\beta=\beta(x)$, $A_i=A_i(x)$.

By {\bf linearity} of the operator $\hat b_{ik}$ (the {\em first} order, \eqref{5}), we have ($c^{-2}\beta Au<<1$):
\begin{equation*}
\nonumber
  f_{ik,lm}=\hat b_{ik}g_{lm}\equiv g_{lm}\hat b_{ik}\ln\chi \simeq2c^{-2}g_{lm}\cdot
 \end{equation*}
\begin{equation}\label{72}
\cdot\left(\beta\hat b_{ik}Au+Au\hat b_{ik}\beta\right)=c^{-2}\beta\,g_{lm}F_{ik},
\end{equation}
where $F_{ik}=\partial A_k/\partial x_i-\partial A_i/\partial x_k$ is Maxwell's electromagnetic field tensor\footnote{Why is $\partial\beta/\partial x^k=0$ in \eqref{68}? The point is that the expression
$\partial f_{i\ \ \ \ \ l}^{\ [k,l]}/\partial x^k$  is a result from action of {\em linear projections} of the gradients $\hat b_i$ and $\hat b_{ik}$. Thet "see"\ only {\em projections of tensor function} like
$Au, f_{ik,lm}u^k$, etc. However, the linearity requires that the expressions $f(x)Au$ are treated as a {\em product} $f(x)$ in projection $Au$, see, e. g. \eqref{72}. That is, $\beta(x)$ behaves like a constant
with respect to $\hat b_i$ and $\hat b_{ik}$! See also Subsection \ref{CovDer}.}.

We immediately notice that the Lorentz contri\-bution  to the acceleration of matter (the geodesics \eqref{17}) has the correct form
\begin{equation}\label{73}
  f_{ik,lm}u^ku^lu^m=\beta c^{-2}F_{ik}u^k.
\end{equation}

The other quantities in Eq. \eqref{70} are:
\begin{eqnarray}
\nonumber
  2f_{i \ \ \ \ l}^{\ [k,l]}=3c^{-2}\beta F_i^{\ k},\quad \xi_{ik}=18c^{-4}\beta^2 F_{il}F_k^{\ l}, \\
\nonumber
  4\pi\tau_{ik}=18\left[-F_{il}F_k^{\ l}+\left(g_{ik}/4\right)\left(2F_{lm}u^m\cdot \right.\right.\\
  \left.\left. \cdot F^{ln}u_n-F_{lm}F^{lm}\right)\right].
\end{eqnarray}

Substitution into Eq. \eqref{70} gives
\begin{equation*}
\frac{\partial\left(\sqrt{-g}\, F_i^{\ k}\right)}{\sqrt{-g}\,\partial x^k}=\frac{4\pi}{3c \Lambda}\left[\left(\frac{\Lambda_0}{2}-\frac{2\Lambda_c}{\varkappa}\right)g_i^{\ l}+\tau_i^{\ l}\right]j_l,
\end{equation*}
since the product $F_i^{\ k}\,\partial \ln\sqrt{-g}/\partial x^k\equiv0$ because the terms belong to different sets, \eqref{5n}-\eqref{5nn}.

From this we can already see that the {\bf coincide\-nce} with the (isotropic) Maxwell equation is achi\-eved if the tensor equality
\begin{equation}\label{74}
  -\frac{1}{3\Lambda}\left[\left(\frac{\Lambda_0}{2}-\frac{2\Lambda_c}{\varkappa}\right)g_i^{\ k}+\tau_i^{\ k}\right]\,j_k=j_i,
\end{equation}
is fulfilled, which has the meaning of the field hypothesis of matter in the general {\em anisotropic} case. (The anisotropy of space as a cause of tensor properties of mass was first demonstrated in pione\-er works of
G. Yu. Bogoslovsky \cite{Bo1973,Bo1992}).

A relationship between the scalars $\mu$ and $\mu_0$, is obtained by projecting \eqref{74} onto the selected direction  $u^i$
\begin{eqnarray*}
\nonumber
 -3\Lambda=\Lambda_0/2-2\Lambda_c/\varkappa-\tau_{ik}u^iu^k,
\end{eqnarray*}
and the choice of the {\em cosmological} "constant"\ as
\begin{equation}
  \label{74t}
  \Lambda_c=(7/4)\varkappa \Lambda_0
\end{equation}
gives a reasonable result for the field hypothesis and the theory being developed ($F_i=F_{ik}u^k$)
\begin{eqnarray}\label{75}
    \mu c^2=\mu_0c^2-\frac{3}{8\pi}\left(2F^2+F_{lm}F^{lm}\right)\sqrt{g_{00}}\, u^0.
\end{eqnarray}

Using the electron as an example, we can estimate the relative value of the field contribution to the mass of charged matter. In the conformal case under consideration, $\rho=\rho_0$, so \eqref{49} gives
\begin{equation}\label{76}
|\beta_e/\sqrt{k}|=\mu_0/\mu_e\simeq 2\cdot 10^{21},\, \mu_e \simeq 5\cdot 10^{-22}\mu_0
\end{equation}
that is, the electron mass is almost completely compensated by the field contribution of its own fields! (The proton has a similar compensation: $\sim 10^{-19}$). And as follows from \eqref{75},  this contribution at
$u^i\simeq (1,\vec v/c), \vec v/c \ll 1$ is the contribution of the magnetic field, $\simeq(-3/4\pi)\,\vec H^2$. Obviously, by varying the magnetic field of the system, one can change both the value and the sign of $\mu$.
(For an experiment on this subject, see Appendix D.)

Thus the conform-Riemann metric \eqref{71t} directly gives Maxwell's vector electrodynamics, which proves Weyl's conjecture on the conformal invari\-ance of gravity in the case of {\em Embedding}.

Finally, the content of this section of the paper can be considered as a special (vector) case of geometrization of electricity. Obviously, the general case will be described by an analog of Eqs. \eqref{69}, \eqref{70},
as a result of extremization of the action of the matter-field system in the {\em tensor} "potential"\ \ $c_{i,kl}$ \eqref{3}.

\section{\textmd{\textmd{CONCLUSION}}}
\label{Concl}
\quad Geometrization of electricity and gravity is possible by replacing of the 4D pseudo-Riemann structureless GR continuum with the {\em inhomogene\-ous} Finsler model of the space-time (MES) -- the
{\em Embedding}.  The model leads to nonlinear (Enstein type) equations of gravity and Maxwell type for electricity, as well as to a model of the physical vacuum: the {\em Embedding} is filled with an isotropic
{\em seed} matter and its fields. The example of a conform-Riemann metric confirms Weil's hypo\-thesis \cite{Weil1918}: it gives the classical vector Maxwell electrodynamics with a natural identification of the
cosmological constant \eqref{74t}, and an explicit expression for the field mass hypothesis: Ed. \eqref{75} directly indicates a {\em negative} electromagnetic field contribution. The latter can explain the small
specific mass of charged particles (electron, proton, etc.): apparently, their seed mass is almost comple\-tely compensated by the contribution of their own fields. in the case of low velocities, the field contribution
reduces to the contribution of the magnetic field, which indicates that the inertial properties of matter can be controlled by external magnetic fields. Evaluation of the required fields is presented in the Appendix D:
the corresponding experiment is possible and necessary, for example, in view of possible technological applications.

The proposed geometrization is remarkable because the quantity $\pm\sqrt{k}$ is a characteristic of the physical vacuum of the {\em Embedding}. This statement can be experimentally verified by mea\-surements
of the electromagnetic "redshift"\ effect, an analog of gravitational redshift. In the special case of an electric field, its calculation is similar to that of the gravitational frequency shift: $\Delta\omega_e/\omega
\simeq \mp 0.861\cdot10^{-21}\cdot\Delta\varphi_e(V)$, where $\varphi_e(V)$ is the elect\-ric field potential in volts, and the sign must be determined experimentally \cite{Nskv2017}. Obviously, the
existence of this effect and its measurements can provide information on both the vacuum and a relic electric charge of the Universe. Finally, if the vacuum of the Universe is similar to the vacuum of the {\em Embedding},
 then it is hypothetically possible to put into correspondence with the quantum "foam"\ vacuum of the Standard Model: $\sqrt{k}\sim\\<|\rho_0/\mu_0|>_{|SM}$. In this relation: cannot the gravi\-ty of the Universe be
 a Casimir effect?

The structure of the MES space, {\em Embedding} of the proper metric sets of matter, is also of fundamental importance for quantum physics. For example, during measurements, the {\em Embedding} metric is
always non-stationary because there is an additional interaction between the device and the object. This leads to a natural {\em inaccuracy} of measurement results.

Since the Maxwell-like equations for {\em contracti\-ons} of $f_{ik,lm}$ are found in an unconventional way, then the final goal of the theory should probably be equations for $f_{ik,lm}$ {\em itself}, which found
in the standard way: a variational least-action procedure for the matter-field system in the anisotropic potential $c_{i,kl}$: $\delta S[g^{ik},c^{i,kl}]_{|g^{ik}=const}=0$. In other words, the present work can be
continued.

\section*{\textmd{Appendix A}}

\subsection*{\textmd{FINSLER PARAMETRIZATIONS: \\ TWO TYPES OF GEOMETRIES}}
\label{AppTeorema}

\quad 1. The basic statement of Finsler geometry is the homogeneity postulate: the metric function $F$ that determines the distance $ds$ between two close points with coordinates $x^i$ and  $x^i+dx^i$ of Finsler
space $F_n$ depends on $x^i$ and $dx^i$ as independent variables and is continuous, smooth, and {\em homogeneous of  $+1$ degree} in the argument $dx$\footnote{To simplify the formulas, we will omit indices
where their absence does not prevent understanding.}
$$
  ds=F(x^i,dx^k),
$$
$$
    F(x,k dx)=k F(x,dx)\quad (k>0).
 \eqno(A.1)
$$

Euler's theorem on homogeneous functions $dx^i\partial F(x,dx)/\partial dx^i=+1\,F(x,dx)$, immediately gives
$$
ds^2=\frac12\frac{\partial^2 F^2(x,dx)}{\partial dx^i\partial dx^k}\, dx^idx^k \quad\Rightarrow
$$
$$
g_{ik}(x,dx)=\frac12\frac{\partial^2 F^2(x,dx)}{\partial dx^i\partial dx^k},
\eqno(A.2)
$$
where $g_{ik}(x,dx)\equiv g_{ik}(x^l,dx^m)$ is interpreted as the metric tensor $F_n$ at point $x^l$ in the direction $dx^m$. Obviously, $g_{ik}(x,dx)$ is a {\em function of $dx$ of homogeneity degree  $0$}.
Euler's theorem also allows us to state that for $C_{ikl}(x,dx)$, a totally symmetric tensor of rank 3,
$$
C_{ikl}(x,dx)=\frac12\frac{\partial g_{ik}(x,dx)}{\partial dx^l}=\frac14\frac{\partial^3 F^2(x,dx)}{\partial dx^i\partial dx^k\partial dx^l},
$$
(a function of homogeneity degree  $-1$ in $dx$), the orthogonality condition holds:
$$
C_{ikl}(x,dx)\,dx^i=C_{ikl}(x,dx)\,dx^k=
$$
$$
=C_{ikl}(x,dx)\,dx^l=0.
\eqno(A.3)
$$

Parametrization of the function $F(x,dx)$ is realized by scalar $t$, a quantity {\em invariant} under arbitrary coordinate transformations $x'^i=x'^i(x^k), \\ (i=1,2,...n)$ (\cite{Rund1959}, \$ 1.2, second paragraph),
and is {\em also} determined by geometric properties of $t$. It is easy to see that, depending on these properties, the "velocity"\ $\dot x\equiv dx/dt$ may {\em be} or may {\em not be} a function of homogeneity degree
$+1$ of $dx$. For example, the admissible {\em non-geometric} parameter "Newtonian time"\ {\em preserves} the homogeneity of the direction argument: $\dot x\equiv dx/dt$ is still a function of homogeneity degree
{\em $+1$} of $dx$.  Convers\-ely, the also admissible {\em natural} parameter $s$ (lenth), a {\em geometric} 1D-extended quantity, changes the anisotropy argument: $\dot x$ is now a function of $dx$ of {\em homogeneity degree $0$}.  Naturally, this variability of $\dot x$ affects the properties of the paramethrized metric function $F(x,\dot x)$, so the parametrization $F(x,dx)$ must always be accompanied by some
{\em additional} condition imposed on its result, on  $F(x,\dot x)$.

\quad 2. {\em Assume} that $F(x,\dot x)$ is a function of {\em homogeneity degree $+1$ } with respect to $\dot x$ ({\em Condi\-tion A} (I.1.8) \cite{Rund1959} -- an {\em additional} condition on homogeneity).
Then for $ds=F(x,\dot x)\,dt$ we imme\-diately have a replica of Eqs.$(A.1)$ -- $(A.3)$, the {\em basic formulas for the standard (homogeneous) version of Finsler geometry \cite{Rund1959} }:
$$
F(x,k\dot x)=kF(x,\dot x)\quad (k>0),
$$
$$
\dot x^i\,\partial F(x,\dot x)/\partial \dot x^i=+1\,F(x,\dot x),
\eqno(A.1')
$$
$$
ds^2=\frac12\frac{\partial^2 F^2(x,\dot x)}{\partial \dot x^i\partial \dot x^k}\, dx^idx^k \quad\Rightarrow
$$
$$
g_{ik}(x,\dot x)=\frac12\frac{\partial^2 F^2(x,\dot x)}{\partial \dot x^i\partial \dot x^k},
\eqno(A.2')
$$
where $g_{ik}(x,\dot x)$ is a function of $\dot x$ of {\em homogeneity degree $0$}, it is the metric tensor of $F_n$ ay the point $x^i$ in the direction $\dot x^k$ (coinciding with $dx^k$). And also, for
$C_{ikl}(x,\dot x)$ (a symmetrical tensor of rank $3$, a function of $\dot x$ of {\em homogeneity degree $-1$}),
$$
C_{ikl}(x,\dot x)=\frac12\frac{\partial g_{kl}(x,\dot x)}{\partial \dot x^i}=\frac14\frac{\partial^3 F^2(x,\dot x)}{\partial \dot x^i\partial \dot x^k\partial \dot x^l},
$$
the orthogonality condition is replicated, it is the {\em most important} property of a standard (homogeneous) version of the geometry:
$$
C_{ikl}(x,\dot x)\,\dot x^i=C_{ikl}(x,\dot x)\,\dot x^k=
$$
$$
=C_{ikl}(x,\dot x)\,\dot x^l=0.
\eqno(A.3')
$$

\quad 3. Let the parameter $t$ be a 1D extended quantity, then $dt$ is function of homogeneity degree {\em $+1$}, but $\dot x^i$ is of {\em degree $0$ in $dx$}\footnote{An example of a metric function is
$ds=f(\dot x)\Phi(x,dx)$, where $f(\dot x)$ is an arbitrary smooth function of $\dot x=dx/ds_{(0)},\, ds_{(0)}^2=g^{(0)}_{ik}dx^idx^k$ and at the same time, it is a function of homogeneity degree $0$
in $dx$, while $\Phi(x,dx)$ is a function of homogeneity degree $+1$ of $dx$.}. Assume that {\em Condition A} (the homogeneity condition) {\em is not valid}: $ds=Q(x,\dot x)dt$, where $Q(x,\dot x)$ is a
smooth function {\em inhomogeneous} in $\dot x$.

Let us prove that the assumed inhomogeneous condition is {\em compatible} with Finsler's homogeneity postulate $(A.1)$. That is, that $F(x,dx)=Q(x,\dot x)dt$ is a function of homogeneity degree $+1$ of $dx$:
$$
dx^i\frac{\partial F(x,dx)}{\partial dx^i}=dx^i\frac{\partial Q(x,\dot x)}{\partial dx^i}\,dt+Q(x,\dot x)\cdot
$$
$$
\cdot\left(dx^i\frac{\partial dt}{\partial dx^i}\right)=dt\,\frac{\partial Q(x,\dot x)}{\partial \dot x^k}\left(dx^i\frac{\partial \dot x^k}{\partial dx^i}\right)+
$$
$$
+(+1)Q(x,\dot x)dt=\left(0\cdot \dot x^k\,\right)dt\frac{\partial Q(x,\dot x)}{\partial \dot x^k}+
$$
$$
+(+1)F(x,dx)=(+1)F(x,dx),\quad \Box.
\eqno(A.1'')
$$

Therefore, Eqs. $(A.1)$ -- $(A.3)$ are also valid in this {\em inhomogeneous} version of the geometry\footnote{For example, $2g_{ik}(x,dx)=Q(x,\dot x)^2\left(\partial^2 (dt)^2/\partial dx^i\partial dx^k\right)$
is a function of homogeneity degree $0$ in $dx$, but from here immediately follows $C_{ikl}(x,dx)\,dx^i=C_{ikl}(x,dx)\,dx^k=C_{ikl}(x,dx)\,dx^l=0$.}. But here links like $(A.1')$ -- $(A.3')$ are {\em absent}
because $F(x,\dot x)=Q(x,\dot x)$ is an {\em inhomogeneous} function! Obviously, this internal freedom of an {\em inhomogeneous} geometry  has fundamental and far-reaching consequences: it is radically different
from the standard homogeneous Finsler version \cite{Rund1959} and can be considered as a certain Finsler generalization of Riemann geometry.

\quad 4. To illustrate the {\em new} Finslerian possibilities when the homogeneity condition {\em Condition A} is violated, we can use the geodesic equation as an example:
$$
\delta\int_{x_1}^{x_2}ds=0, \quad ds^2=g_{ik}(x,\dot x)dx^idx^k\quad \Rightarrow
\eqno(A.4)
$$
$$
\ddot{x_i}-\left(\frac12\frac{\partial g_{kl}(x,\dot x)}{\partial x^i}+f_{im,kl}\dot x^m\right)\dot x^k\dot x^l-\frac{\ddot{s}}{\dot s}\,\dot x_i+
$$
$$
+\left(C_{mkl}\frac{\partial ln\dot s}{\partial x^i}-C_{ikl}\frac{\partial ln\dot s}{\partial x^m}\right)\dot x^k\dot x^l\dot x^m=0,
\eqno(A.5)
$$
where
$$
f_{im,kl}=\frac{\partial C_{mkl}}{\partial x^i}-\frac{\partial C_{ikl}}{\partial x^m},
$$
$$
\frac{\partial ln\dot s}{\partial x^i}=\frac12\frac{\partial g_{np}(x,\dot x)}{\partial x^i}\frac{\dot x^n\dot x^p}{\dot x^2},
$$
$$
\frac{\ddot s}{\dot s}=\frac{1}{\dot x^2}\left[\ddot x^k \dot x_k+\left(\frac12\frac{\partial g_{kl}(x,\dot x)}{\partial x^m}\dot x^m+C_{mkl}\ddot x^m\right)\dot x^k\dot x^l\right].
$$

This immediately shows that for the standard {\em homogeneous} Finsler geometry, when {\em Condition A is satisfied}, i.e. the orthogonality $(A.3')$ also holds, the Loretzian term $f_{im,kl}\dot x^m\dot x^k\dot x^l$,
as well as the last two terms, automatically fall out from the geodesic (are zero). But if we choose $t=s$, то $(A.5)$ becomes a geodesic equation with only an anisotropic Riemann term:
$$
\frac{du_i}{ds}-\frac12\frac{\partial g_{kl}(x,u)}{\partial x^i}u^ku^l=0,\quad u_i=\frac{dx_i}{ds}.
\eqno(A.5')
$$

In {\em inhomogeneous} Finsler geometry, when {\em Con\-dition A is not satisfied}, we have both anisotropic Riemann and Lorentzian terms, see \eqref{4}
$$
\frac{du_i}{ds}- \left(\frac12\frac{\partial g_{kl}(x,u)}{\partial x^i}+f_{im,kl}(x,u)u^m\right)u^ku^l=0,
\eqno(A.5'')
$$
$$
f_{im,kl}(x,u)=\frac{\partial c_{m,kl}(x,u)}{\partial x^i}-\frac{\partial c_{i,kl}(x,u)}{\partial x^m},
$$
$$
c_{i,kl}(x,u)=\frac12\frac{\partial g_{kl}(x,u)}{\partial u^i},
$$
where $c_{i,kl}(x,u)$ is the tensor electromagnetic potential, symmetric with respect to the last indi\-ces.

That is, the {\em new} Finslerian possibilities reduce to the emergence of a Lorentzian term in the geodesic equation (equivalent to the possibility of geometrization of electrodynamics in $F_n$) and a lot of new physical
content of the geometry.

In conclusion, we can say that the standard (homogeneous) Finsler geometry has two funda\-mental {\em non-relativistic} features: it admits a Newtonian parametrization (Newtonian time)\footnote{This non-relativistic
nature can be proved: the requirement of $+1$ degree of homogeneity of $F(x,\dot x)$ in  $\dot x$ (Condition A  (I.1.8) in \cite{Rund1959}) with the postulate that "$F(x,dx)$ is a function of homogeneity degree $+1$
in $dx$"\  means that $\dot x=dx/dt$ can be {\bf only} $+1$ degree in $dx$! Therefore, $dt$ {\bf cannot be} 1D-extended quantity: only non-geometric parameters are allowed, such as Newtonian time. Therefore, the
  standard (homogeneous) Finsler geometry is {\bf non-relativistic}:  Condition A  excludes other options for it. $\Box$.},  and its geodesic {\em lacks} a Lorentzian term. A good example of this is given by 3D space.

Thus, standard Finsler geometry {\em cannot be used in relativistic physics}. One of the indirect indications of such a harsh conclusion is found in G. Yu. Bogoslovsky's paper: "\ ... the general theory of Finsler spaces
suggests the existence of not one (as in the  Riemann case), but three fourth-rank tensors describing the curvature. ... This circumstance creates certain difficulties on the way of formal generalization of the Hilbert-Einstein equations using the Finsler apparatus"\ , \cite{Bo1992}.

Conversely, the {\em inhomogeneous} Finsler geomet\-ry {\em can be used in relativistic physics} and is applica\-ble to geometrization of classical electromagne\-tism, see, for example, Section 2. The main feature of the
class of {\em relativistic} Finsler geometries is the {\em inhomogeneity} of the metric function $F(x,\dot x)$ in $\dot x$.

\section*{\textmd{Appendix B}}

\subsection*{\textmd{BOGOSLOVSKY'S ANISOTROPIC METRIC}}
\label{Bogoslovsky}
\quad A comparison of the geometry of this paper with Bogoslovsky's conform-flat anisotropic geometry \cite{Bo1973,Bo1992,Bo1996,Bo2012} allows us to make several obser\-vations. First, Bogoslovsky's metric
function
$$
ds=\left[\frac{(dx_0-\vec{\nu}d\vec{x})^2}{dx_0^2-d\vec{x}^{\, 2}}\right]^{r/2}\sqrt{dx_0^2-d\vec{x}^{\, 2}},
$$
where $r=const$, $\vec{\nu}=const$ is a fixed unit vector,  satisfies Finsler's homogeneity postulate for the metric function $F(x,dx)$ (A.1). Second, the corresponding metric can be represented as
$$
g_{ik}(u)=\left(g^{(0)}_{lm}\nu^l u^m\right)^r g^{(0)}_{ik},
\eqno(B.1)
$$
where  $i,k,...=0,1,2,3$, $\nu_i=(1,-\vec{\nu})$ is a null vector ($1-\vec{\nu}^2=0$),  $u^i=dx^i/ds^{(0)}$ and $ds^{(0)}=+\sqrt{g^{(0)}_{ik}dx^idx^k}$ is the Minkowski interval (flat tangent 4D space
with the metric tensor $g^{(0)}_{ik}$). The vector $u$ should be chosen as the velocity $\dot x$ due to its {\em general covariance}: $ds$ is a scalar invariant under {\em arbitrary} coordinate transforma\-tions
$x'^i=x'^i(x^k),\ (i=1,2,...n)$.

The expression (B.1) shows that the Bogoslov\-sky's metric {\em is not} a function of {\em zero degree of homogeneity} with respect to $u$, that is, it does not belong to the standard  Finsler metric. It belongs to the class
of relativistic Finsler metrics: the Lorentzian term of the corresponding geodesic is zero {\em only} by virtue of $\vec{\nu}=const$, and the physical meaning of space is an anisotropic analog of Minkowski space with an
abstract anisotropy vector $\vec{\nu}=const$.

In the more general case $\nu_i(x)$, the geodesic has a Lorentzian term that is not equal to zero, and at $r=2$ the vector itself can be interpreted as $\nu_i(x)\sim A_i$, where $A_i$ is the electromagnetic field potential
of Bogoslovsky space. For example, $c^2\nu_i(x)=\pm\sqrt{k} A_i(x)$.

In the case of electric neutrality of the metric (B.1), the vector $\nu_i(x)$ can probably be interpreted as the angular velocity for models of the Universe with rotation \cite{Panov1984}.

\section*{\textmd{Appendix C}}

\subsection*{\textmd{DIFFERENTIATION IN $u^i$}}
\label{Dif u}
\quad The derivative in the vector $u^i$, often used  in this paper, is nontrivial because of the normaliza\-tion $g_{ik}u^iu^k-1=0$: only three of its four components are actually independent. Naturally, when using the
derivative $\partial/\partial u^i$, this fact must somehow be taken into account.

In our opinion, there is still one universal way to solve the problem of differentiation with respect to the unit vector $u^i$: replacing the gradient $\partial/\partial u^i$ with the variational derivative $\delta/\delta u^i$
using the method of indefinite Lagrange multipliers (analo\-gous to the variational problem for a functional with connections):
$$
\frac{\partial f(u^k)}{\partial u^i}\Rightarrow \frac{\delta f^*(u^k)}{\delta u^i}\equiv \frac{\partial \left[f(u^k)+\lambda(u_iu^i-1)\right]}{\partial u^i}
$$
$$
=\left(\frac{\partial}{\partial u^i}+\lambda^* u_i\right)f(u^k)\Rightarrow \frac{\delta}{\delta u^i}=\frac{\partial}{\partial u^i}+\lambda^* u_i,
\eqno(C.1)
$$
where $\lambda^*=\lambda/f(u^k)\neq \lambda^*(u^i)$ is an indefinite  Lagrangian multiplier, while the derivative $\partial/\partial u^i$ is taken as with respect to a usual vector with {\em independent components}.

Let us find $\lambda^*$. We choose as $f(u^k)$ the norm $g_{kl}(x,u)u^ku^l$=1:
$$
0=\frac{\delta (g_{kl}u^ku^l)}{\delta u^i}=u^ku^l\frac{\delta g_{kl}}{\delta u^i}+2g_{kl}u^l\frac{\delta u^k}{\delta u^i}=
$$
where $\delta g_{kl}/\delta u^i\equiv c_{i,kl}$ is a derivative in the normalised vector (!), and to calculate $\delta u^k/\delta u^i$ we use the result (C.1). Then, continuing, we have
$$
=2c_{i,kl}u^ku^l+u_k\left(\frac{\partial}{\partial u^i}+\lambda^*u_i\right)u^k=
$$
$$
=2c_{i,kl}u^ku^l+u_i+\lambda^*u_i\Rightarrow \lambda^*=1.
\eqno(C.2)
$$

Thus a derivative with respect to to the unit vector $u^i$ should be taken as variational one:
$$
\frac{\delta}{\delta u^i}=\frac{\partial}{\partial u^i}+u_i,
\eqno(C.3)
$$
moreover, the vector $u^i$ in the gradient $\partial/\partial u^i$ {\bf (as in $\hat b_i$)} is already understood as {\bf non-normalized} one (all its components are independent).

Returning to the geodesic equation in the {\em Embedding} \eqref{4}, \eqref{16}, it becomes clear that it is valid if the velocity field vector $u^i$ is orthogonal to its 4D curl:
$$
\left(\partial u_k/\partial x^i-\partial u_i/\partial x^k\right)u^i=0.
\eqno(C.4)
$$

{\bf Proof.} In the derivation of the geodesic equa\-tion, before Eq. \eqref{3} we have ($u^i=u^i(s)$):
$$
0=ds\left\{\delta x^i[...]-\delta x^i\frac{d\vartheta_i}{ds}+\delta_ x s\frac{d(\vartheta u)}{ds}\right\},
$$
$$
\vartheta_i=\frac{u^ku^l}{2}\frac{\delta g_{kl}}{\delta u^i}.
\eqno(C.5)
$$
Switching to field operators gives
$$
0=ds\left\{\delta x^i[...]-\delta x^iu^k\frac{\partial \vartheta_i}{\partial x^k}+\delta x^i \frac{\partial(\vartheta_k u^k)}{\partial x^i}\right\}
$$
$$
\equiv\delta x^i ds\left\{[...]-u^k\left(\frac{\partial \vartheta_i}{\partial x^k}-\frac{\partial \vartheta_k}{\partial x^i}\right)\right\},
$$
since $u^i=u^i(s)$. For the same reason,
$$
0=\delta x^ids\left\{[...]+u^k\left(\frac{\partial \vartheta_k}{\partial x^i}-\frac{\partial \vartheta_i}{\partial x^k}\right)\right\}=\delta x^ids
$$
$$
\cdot\left\{[...]+\frac{u^ku^lu^m}{2}\left(\frac{\partial}{\partial x^i}\frac{\delta}{\delta u^k}-\frac{\partial}{\partial x^k}\frac{\delta}{\delta u^i}\right)g_{lm}\right\},
$$
see also (C.5). Then, in accord with (C.3), we have
$$
0=\delta x^ids\left\{[...]+\frac{u^ku^lu^m}{2}\left[\frac{\partial}{\partial x^i}\left(\frac{\partial}{\partial u^k}+u_k\right) \right. \right.
$$
$$
\left. \left.-\frac{\partial}{\partial x^k}\left(\frac{\partial}{\partial u^i}+u_i\right)\right]g_{lm}\right\}
$$
$$
=\delta x^ids\left\{\frac{du_i}{ds}-\frac12u^ku^l\left(\frac{\partial}{\partial x^i}+2u^m\frac{\partial^2}{\partial x^{[i}\partial u^{m]}}\right)g_{kl}\right.
$$
$$
\left.-\frac12u^k\left(\frac{\partial u_k}{\partial x^i}-\frac{\partial u_i}{\partial x^k}\right)\right\},
$$
more over, in the last term formally $u_i=u_i(x,u^k=const)\equiv u_i(x)$ is the Eulerian velocity field. That is, the motion of the continuous medium under consideration is such that the velocity field is orthogonal
to its curl:
$$
u_{[k,i]}u^k=0,\quad \Box.
\eqno(C.6)
$$

\section*{\textmd{Appendix D}}

\subsection*{\textmd{AN EXPERIMENT FOR TESTING \\THE FIELD HYPOTHESIS}}
\label{ExpFH}
\quad The result \eqref{75} is very important for application purposes: the real mass of a system can be {\em reduced} by changing a local electromagnetic field. It is interesting to estimate the fields required to
comp\-letely compensate the mass of a proton and electron.

According to \eqref{76}, for uncompensated (real) mass of a particle we can write
$$
\mu c^2\simeq(3/4\pi)\vec H_0^2,
\eqno(D.1)
$$
where $\vec H_0$ is the magnetic field that reduces the mass of the particle at rest ti zero. then, assuming for estimation purposes, that the particle is a homogeneous ball with a radius equal to its Comp\-ton length
$\lambda_c$,
$$
\vec H_0^2\simeq m c^2/\lambda_c^3\, \sqrt{\overline{\gamma}},
\eqno(D.2)
$$
where $\overline{\gamma}=\overline{\det[\gamma_{\alpha\beta}]}$ is the average value of the determinant of the 3D tensor $\gamma_{\alpha\beta}$ over the particle volume.

Using the standard data for the proton and the electron, it is not difficult to find
$$
H_{0p}\approx 8,1\cdot 10^{17}Oe/(\overline{\gamma_p})^{1/4},
$$
$$
H_{0e}\approx 2,4\cdot 10^{11}Oe/(\overline{\gamma_e})^{1/4}.
\eqno(D.3)
$$

And since $\gamma$ is mainly determined by the self-interacting, we should expect very large values for the denominator. It is for this reason that $H_0$ may be a reasonably achievable quantity, and the mass
variation experiment may be {\em feasible}...

\renewcommand{\refname}{

%Keywords: {\it geometrization, \ electrodynamics, \ gravitation, \ Model of Embedded Spaces}

\end{document}